\documentclass[showpacs,amsmath,amssymb,aps,twocolumn,longbibliography,pra]{revtex4-1}
\usepackage{graphicx}
\usepackage{soul}\usepackage[colorlinks=true,citecolor=blue,linkcolor=magenta]{hyperref}
\usepackage[usenames]{color}
\usepackage{relsize}
\usepackage[english]{babel}
\usepackage{enumerate}
\usepackage{pifont}
\usepackage{amsfonts}

\newcommand{\ket}[1]{\left| #1 \right>} 
\newcommand{\bra}[1]{\left< #1 \right|} 

\newcommand {\grsim} {\ {\raise-.5ex\hbox{$\buildrel>\over\sim$}}\ }
\newcommand {\lessim} {\ {\raise-.5ex\hbox{$\buildrel<\over\sim$}}\ }

\begin{document}

\title{Realization of an anomalous Floquet topological system with ultracold atoms}

\author{Karen~Wintersperger$^{1,2}$, Christoph~Braun$^{1,2,3}$, F.~Nur~\"Unal$^{4,5}$, Andr\'{e}~Eckardt$^{4,6}$, Marco~Di~Liberto$^{7}$, Nathan~Goldman$^{7}$, Immanuel~Bloch$^{1,2,3}$, Monika~Aidelsburger$^{1,2}$}

\affiliation{$^{1}$\,Fakult\"at f\"ur Physik, Ludwig-Maximilians-Universit\"at M\"unchen, Schellingstra{\ss}e 4, 80799 M\"unchen, Germany}
\affiliation{$^{2}$\,Munich Center for Quantum Science and Technology (MCQST), Schellingstra{\ss}e 4, 80799 M\"unchen, Germany}
\affiliation{$^{3}$\,Max-Planck-Institut f\"ur Quantenoptik, Hans-Kopfermann-Stra{\ss}e 1, 85748 Garching, Germany}
\affiliation{$^{4}$\,Max-Planck-Institut f\"ur Physik komplexer Systeme, N\"othnitzer Stra{\ss}e 38, 01187 Dresden, Germany}
\affiliation{$^{5}$\,T.C.M. Group, Cavendish Laboratory, 19 JJ Thomson Avenue, Cambridge, CB3 0HE, UK}
\affiliation{$^{6}$\,Institut f\"ur Theoretische Physik, Technische Universit\"at Berlin, Hardenbergstra{\ss}e 36, 10623 Berlin, Germany}
\affiliation{$^{7}$\,Center for Nonlinear Phenomena and Complex Systems, Universit\'{e}  Libre de Bruxelles, CP 231, Campus Plaine, B-1050 Brussels Belgium}


\maketitle

\textbf{
Coherent control via periodic modulation, also known as Floquet engineering, has emerged as a powerful experimental method for the realization of novel quantum systems with exotic properties. 
In particular, it has been employed to study topological phenomena in a variety of different platforms. In driven systems, the topological properties of the quasienergy bands can often be determined by standard topological invariants, such as Chern numbers, which are commonly used in static systems. 
However, due to the periodic nature of the quasienergy spectrum, this topological description is incomplete and new invariants are required to fully capture the topological properties of these driven settings. 
Most prominently, there exist two-dimensional anomalous Floquet systems that exhibit robust chiral edge modes, despite all Chern numbers are equal to zero.  
Here, we realize such a system with bosonic atoms in a periodically-driven honeycomb lattice and infer the complete set of topological invariants from energy gap measurements and local Hall deflections.
}


Floquet engineering~\cite{goldman_periodically_2014,bukov_universal_2015,eckardt_colloquium_2017} has found widespread applications for the realization of out-of-equilibrium many-body systems with novel properties in systems of ultracold atoms~\cite{struck_quantum_2011,aidelsburger_experimental_2011}, photonics~\cite{rechtsman_photonic_2013,hafezi_imaging_2013}, superconducting qubits~\cite{roushan_chiral_2017} and graphene~\cite{mciver_light-induced_2020}. 
It plays a key role in many successful realizations of artificial gauge fields and topological lattice models~\cite{aidelsburger_artificial_2018,cooper_topological_2019}, including the paradigmatic Harper-Hofstadter~\cite{aidelsburger_realization_2013,miyake_realizing_2013} and Haldane model~\cite{rechtsman_photonic_2013,jotzu_experimental_2014} and more recently the generation of non-trivial Chern bands in a 2D optical Raman lattice~\cite{wu_realization_2016}. 
The topological properties of non-interacting two-dimensional (2D) lattice models without additional symmetries are well understood by a set of Chern numbers $\mathcal{C}^\mu$ -- a 2D invariant defined as the integral of the Berry curvature $\Omega^\mu(\mathbf{q})$ in quasimomentum space for the $\mu$th energy band: $\mathcal{C}^\mu=\frac{1}{2\pi}\int_{\text{BZ}}\Omega^\mu(\mathbf{q})\,\text{d}^2q$~\cite{thouless_quantized_1982,xiao_berry_2010}, where BZ denotes the Brillouin zone. 
In cold-atom systems a number of experimental techniques has been developed to determine the geometric properties of Floquet quasienergy bands in analogy to their static counterparts~\cite{jotzu_experimental_2014,aidelsburger_measuring_2015,mittal_measurement_2016,flaschner_experimental_2016,tarnowski_measuring_2019,asteria_measuring_2019}. The properties of chiral edge modes on the other hand have been mostly studied with photonic platforms~\cite{rechtsman_photonic_2013,hafezi_imaging_2013}.

\begin{figure}[!tb]
\includegraphics{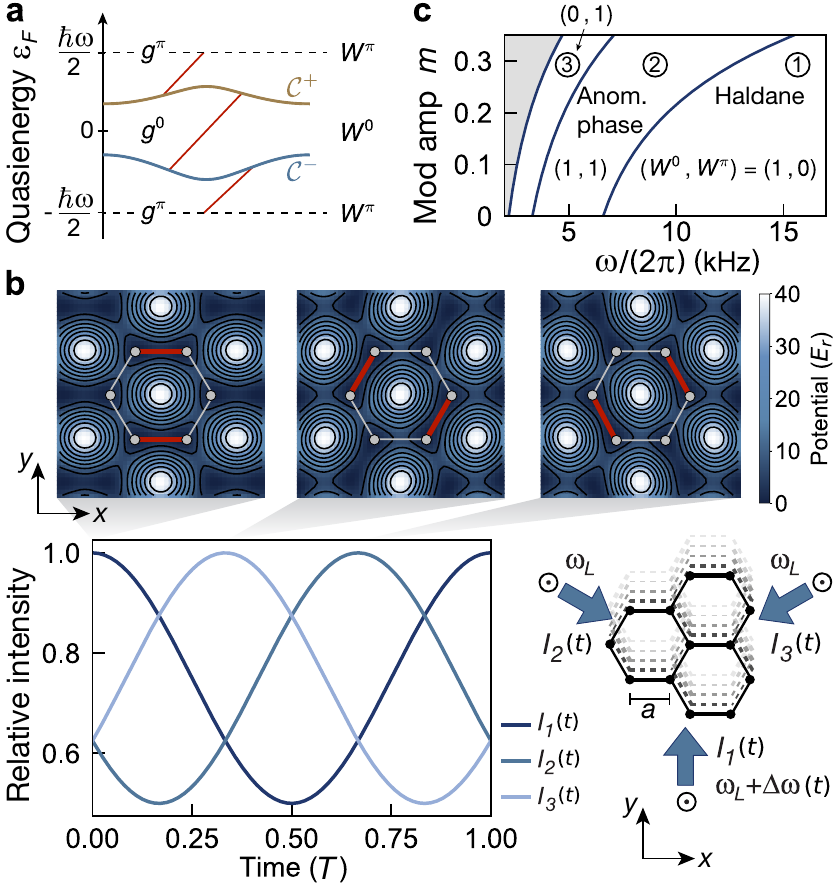}
\vspace{-0.cm} \caption{Schematics of the periodically-modulated lattice, its Floquet quasienergy spectrum and the topological phase diagram. \textbf{a}. Exemplary Floquet spectrum (reduced zone scheme) with topological invariants: Chern numbers $\mathcal{C}^{\pm}$ and Winding numbers $W^{j}$, which determine the number and chirality of edge modes (red lines) in gap $g^{j}$, $j\in\{ 0,\pi\}$. 
\textbf{b}. Periodic modulation of the laser intensities $I_i(t)$, $i=\{ 1,2,3\}$ and the resulting real-space potential over one period of the driving $T=2\pi/\omega$, with $\omega$ the modulation frequency. Red lines indicate larger tunnel couplings.
Right panel: Illustration of the three interfering laser beams with frequency $\omega_L$ and the modulated lattice with constant $a=284\,$nm. Lattice acceleration is realized with additional time-dependent detunings $\Delta\omega (t)$ (Methods). 
\textbf{c}. Topological phase diagram.  
Phase boundaries are obtained from a Floquet band structure calculation (Supplementary Information). 
The gray shaded area contains additional phases not discussed in this work.}
\label{Fig_1}
\end{figure}

For static 2D systems, such as Chern insulators, there is a direct correspondence between the Chern number of the bulk band and the net number of topologically protected 1D edge modes at the boundaries of the sample, known as \textit{bulk-edge correspondence}~\cite{hatsugai_chern_1993,qi_general_2006}. 
Remarkably, this correspondence survives for certain classes of periodically-driven systems in the high-frequency limit, where the modulation frequency is the largest energy scale in the system. 
In general, however, the bulk-edge correspondence is modified and knowledge about the Chern numbers is not sufficient to determine the number and chirality of chiral edge modes in these driven settings~\cite{kitagawa_topological_2010,rudner_anomalous_2013,nathan_topological_2015}. 
Instead this information can be obtained from a new bulk topological invariant, the winding number $W$ (Fig.~\ref{Fig_1}a), characterizing the topology of the quasienergy gaps~\cite{rudner_anomalous_2013}, which depends on the full time evolution during one period of the drive (i.e. the micromotion).
 
In periodically driven systems, the quasienergy $\varepsilon_F$ is only defined up to integer multiples of the driving energy quantum $\hbar \omega$, with angular modulation frequency $\omega$ and reduced Planck's constant $\hbar$. 
Hence, the edge-state dispersion can leave the spectrum from the top and re-enter from below. An exemplary spectrum is illustrated in Fig.~\ref{Fig_1}a, which shows a Floquet quasienergy spectrum in the reduced zone scheme with $-\hbar\omega/2 < \varepsilon_F < \hbar\omega/2$ being the first Floquet Brillouin zone (FBZ) [there are infinitely many copies spaced by $\hbar\omega$]. 
Here, the presence of the edge mode is the result of a non-trivial winding of the quasienergy spectrum itself. This implies that there is an anomalous Floquet topological phase~\cite{rudner_floquet_2019}, where topological edge modes are present, although the Chern number of the energy bands is zero ($\mathcal{C}^{\pm}=0$); here $\mathcal{C}^-$ and $\mathcal{C}^+$ denote the Chern number of the lower and upper quasienergy band in a two-band model. 

Anomalous Floquet systems are genuine time-dependent settings without any static counterpart. 
In particular, these anomalous edge modes exhibit a remarkable robustness that can even exceed those of conventional quantum Hall systems~\cite{nathan_anomalous_2019}.
 Anomalous edge modes have been observed in photonic~\cite{kitagawa_observation_2012,hu_measurement_2015,maczewsky_observation_2017,mukherjee_experimental_2017} and phononic experiments~\cite{peng_experimental_2016}. 
However, a complete experimental characterization of the topological properties of Floquet systems is still lacking. 
Here we report on experimental results obtained with ultracold bosonic atoms in a periodically-modulated honeycomb lattice, where we use a combination of energy gap~\cite{zenesini_observation_2010,kling_atomic_2010} and local Hall drift measurements~\cite{jotzu_experimental_2014,aidelsburger_measuring_2015} in order to reveal the full set of bulk topological invariants.

In honeycomb lattices anomalous Floquet phases can be generated via step-wise periodic modulation of the tunnel couplings~\cite{kitagawa_topological_2010}, using step-wise linear phase shaking~\cite{quelle_driving_2017} or circular phase shaking near resonant with a sublattice energy offset~\cite{unal_how_2019}. 
Here, we employ a continuous analog of the step-wise modulation protocol proposed in Ref.~\cite{kitagawa_topological_2010} using amplitude modulation. 
This is realized by sinusoidal modulation of the laser intensities (Fig.~\ref{Fig_1}b): $I_i(t)=I_0(1- m + m\, \text{cos}(\omega t + \phi_i))$, where $m$ is the relative modulation amplitude and $\phi_i=\frac{2\pi}{3}\times(i-1)$ denotes the modulation phase for the three laser beams, $i=\{1,2,3\}$. 
This time-dependent lattice model exhibits a rich topological phase diagram as a function of the modulation parameters (Fig.~\ref{Fig_1}c). 
We study the geometric properties of the quasienergy spectrum in the three most robust phases: \ding{172} The topological Haldane phase with $\mathcal{C}^{\pm}=\mp1$, \ding{173} an anomalous phase with trivial Chern bands $\mathcal{C}^{\pm}=0$, but chiral edge modes and \ding{174} a Haldane-like topological phase with $\mathcal{C}^{\pm}=\pm1$, where the chiral edge modes are located between Floquet zones. 
Generally, the Chern number of a certain energy band is determined by the difference between the net number of edge modes leaving the band at the top and entering from below, i.e., $\mathcal{C}^\pm=\mp(W^0-W^\pi)$, where $W^{j}$ characterizes the net number of edge modes in the energy gap $g^{j}$, $j\in\{ 0,\pi\}$.
This set of winding numbers uniquely defines the Chern numbers of the bulk bands, however, the opposite only holds for static systems.
In this work we deduce the value of the bulk winding numbers by tracking the topological charges associated with each energy-gap-closing point~\cite{unal_how_2019} that occurs at the topological phase transition, providing a full classification of the Floquet topological phase diagram of our model (Fig.~\ref{Fig_1}c). In addition, we calculated the Floquet quasienergy spectrum for a semi-infinite system using an approximate tight-binding model, which directly reveals the edge modes in the respective quasienergy gaps (Supplementary Information).

The experimental setup consists of a Bose-Einstein-condensate (BEC) of $^{39}$K atoms loaded into an optical honeycomb lattice, that is created by interfering three $s$-polarized laser beams with $\lambda_L=736.8\,\text{nm}$ at relative angles of $120^{\circ}$ (Fig.~\ref{Fig_1}b). 
Additional harmonic confinement is provided by a crossed dipole trap and a third, vertical dipole beam, all at $1064\,\text{nm}$, with total trapping frequency $\omega_r = 2\pi \times 27.0 (4)\, \text{Hz}$ in the $xy$-plane and $\omega_z \approx 2\pi \times 200 \, \text{Hz}$. Using a Feshbach resonance at $403.4(7)\,\text{G}$, we generate a nearly non-interacting cloud with a scattering length of $a_s = 6.35\,a_0$. 
The initial state for all measurements described in the following is a condensate in the lowest energy eigenstate at zero quasi-momentum ($\Gamma$-point) prepared in a honeycomb lattice with in-plane depth $V=6.00(5)\,E_r$, where $E_r=\frac{\hbar^2 k_L^2}{2m_{\text{K}}}=  h \times 9.43\, \text{kHz}$ is the recoil energy, $k_L=2 \pi / \lambda_L$ and $m_{\text{K}}$ is the mass of an atom.


\begin{figure*}[!htb]
\includegraphics[width=\textwidth]{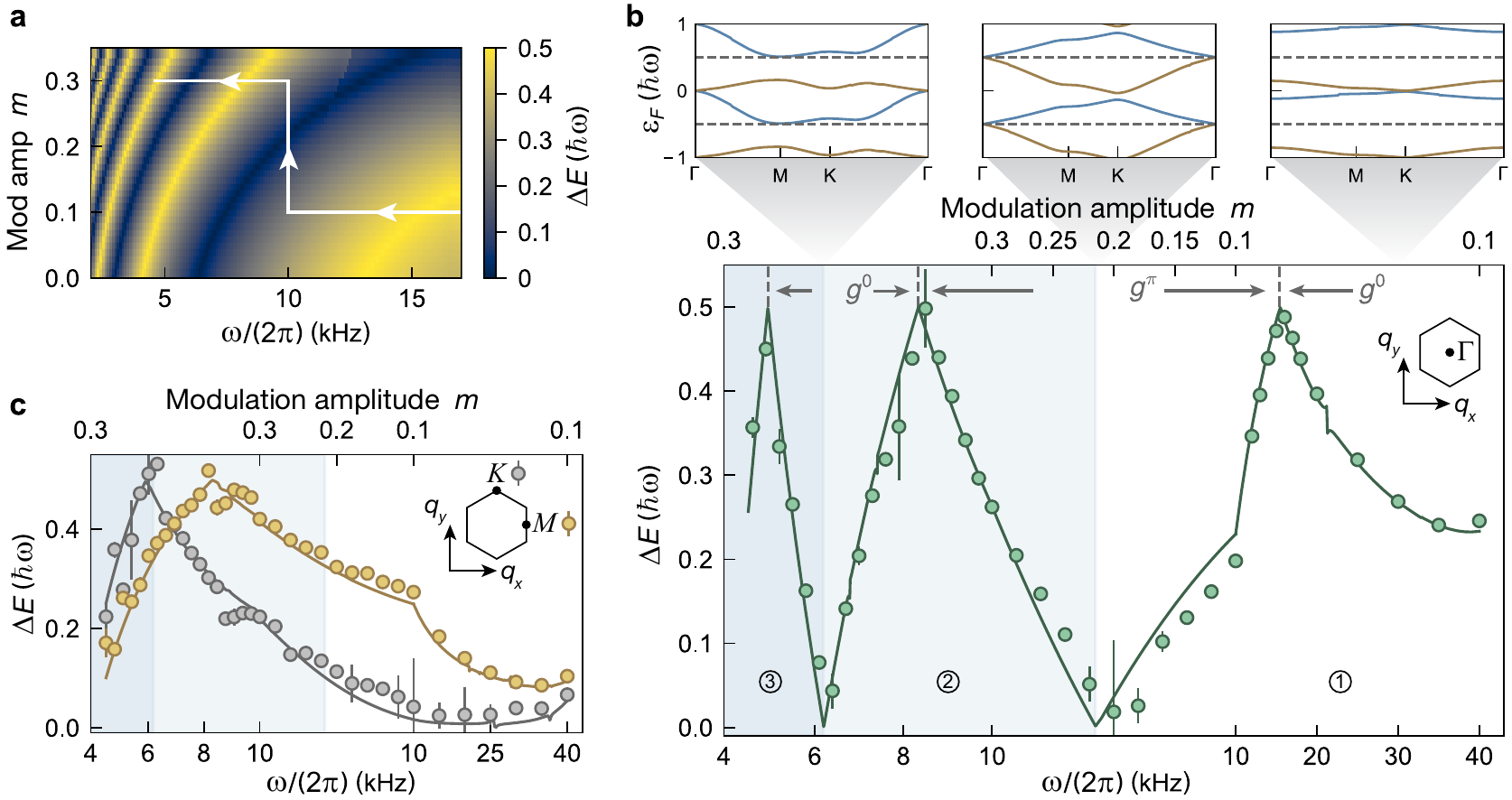}
\vspace{-0.cm} \caption{Energy gaps at $\Gamma$, $K$ and $M$ and energy bands in the different topological phases shown in the extended zone scheme.
\textbf{a}. Minimal energy gap $\Delta E=\min(g^0,g^\pi)$ at $\Gamma$ as a function of modulation frequency and amplitude. 
The white arrows mark the parameter scan used for the measurements in \textbf{b} and \textbf{c}. 
\textbf{b}. Measured energy gaps (points) at $\Gamma$ with $F a/h = 4086 \,\text{Hz}$ and theoretical values (solid line) obtained from a band structure calculation including the six lowest energy bands of the modulated lattice (Supplementary Information). 
Error bars denote fitting errors; every data point represents a St\"uckelberg interferometry measurement with 23 points each being averaged over 3-4 single experimental realizations. 
The blue shaded areas indicate the different topological phases, deduced from the gap-closing points. 
Upper panels: Lowest two energy bands in the extended zone scheme along the high-symmetry path in the first BZ calculated at $\omega/(2 \pi) = 30\,\text{kHz}$, $m=0.1$; $\omega/(2 \pi) = 10\,\text{kHz}$, $m=0.215$; $\omega/(2 \pi) = 6.2\,\text{kHz}$, $m=0.3$ from right to left. 
\textbf{c}. Measured and calculated energy gaps at $K$ and $M$ with errorbars similar to \textbf{b}.}
\label{Fig_2}
\end{figure*}

In a first set of measurements we locate the phase boundaries shown in Fig.~\ref{Fig_1}c by probing the quasienergy gaps. 
This uniquely determines the topological phase transition points, since the topology of the bands can only change via a gap-closing in the spectrum. 
We resolve the energy gap locally as a function of quasimomentum using St\"uckelberg interferometry~\cite{zenesini_observation_2010,kling_atomic_2010} (Methods):
The quasimomentum of the condensate is changed non-adiabatically using lattice acceleration, which results in a coherent superposition of population in the first and second band.
Holding at a specific final quasimomentum $\mathbf{q}$ and subsequently driving back with the same force, produces oscillations of the relative band population with a frequency $\Delta E(\mathbf{q})/\hbar$, which can be measured using bandmapping~\cite{greiner_exploring_2001}. 
Note that in principle there are two different energy gaps $g^0$ and $g^\pi$ that could be probed using this method. However, since we probe the system at a fixed quasimomentum $\mathbf{q}$ at stroboscopic times, we always measure the minimal quasienergy gap $\min(g^0(\mathbf{q}),g^\pi(\mathbf{q}))$ (Methods), which can be chosen to lie in the interval $[0,\hbar\omega/2 ]$ as discussed in~\cite{unal_how_2019}.

Figure~\ref{Fig_2} shows experimental data for various modulation parameters along the path illustrated by the white arrows in Fig.~\ref{Fig_2}a, which covers all three topological phases (Fig.~\ref{Fig_1}c). 
The experimental results of the energy gaps at the high-symmetry points are shown in Fig.~\ref{Fig_2}b for the $\Gamma$-point and in Fig.~\ref{Fig_2}c for the $M$- and $K$-points. Probing the high-symmetry points is sufficient in two dimensions to detect gap-closing points in a honeycomb lattice with our modulation scheme~\cite{bouhon_wilson_2019}.
The obtained data is in excellent agreement with an ab-initio Floquet-bandstructure calculation, which includes the first six bands. While the two lowest quasienergy bands determine the nature of the topological phase diagram, their shape is modified due to hybridization between the $s$- and $p$-bands during one modulation period, which needs to be taken into account for quantitative comparisons (Supplementary Information).
This parameter scan allows us to identify the phase transitions from gap-closing points at $\Gamma$. 
The energy gaps at the $M$- and $K$-points remain finite for all modulation parameters (Fig.~\ref{Fig_2}c). 

Since we always determine the minimal gap $\min(g^0(\mathbf{q}),g^\pi(\mathbf{q}))$, we can use this data to unambiguously identify whether the gap closes within or between Floquet zones~\cite{unal_how_2019}. 
In the high-frequency limit, where $\hbar\omega$ is much larger than any other energy scale, one always measures the gap around zero energy $g^0$, because $g^\pi \gg g^0$ for all quasimomenta. 
Following the parameter scan shown in Fig.~\ref{Fig_2}b, the energy gap $g^0 (\Gamma)$ increases for smaller modulation frequencies until $g^\pi(\Gamma)=g^0(\Gamma)=\hbar\omega/2$ at the first cusp, whereupon we measure the energy gap between Floquet zones, $g^\pi(\Gamma)$. 
We then continue to probe $g^\pi(\Gamma)$, until a second cusp appears, indicating that $g^{0}(\Gamma)$ is now the smaller gap. 
From this we conclude that the first phase transition between phases \ding{172} and \ding{173} occurs via a band touching at the $\Gamma$-point between Floquet zones, while the second one between \ding{173} and \ding{174} appears via a gap-closing at $\Gamma$ around zero energy (upper panels in Fig.~\ref{Fig_2}b).
For $\omega \rightarrow 0$, additional phase transitions occur, which are not discussed in this work.

\begin{figure*}[htb]
\includegraphics{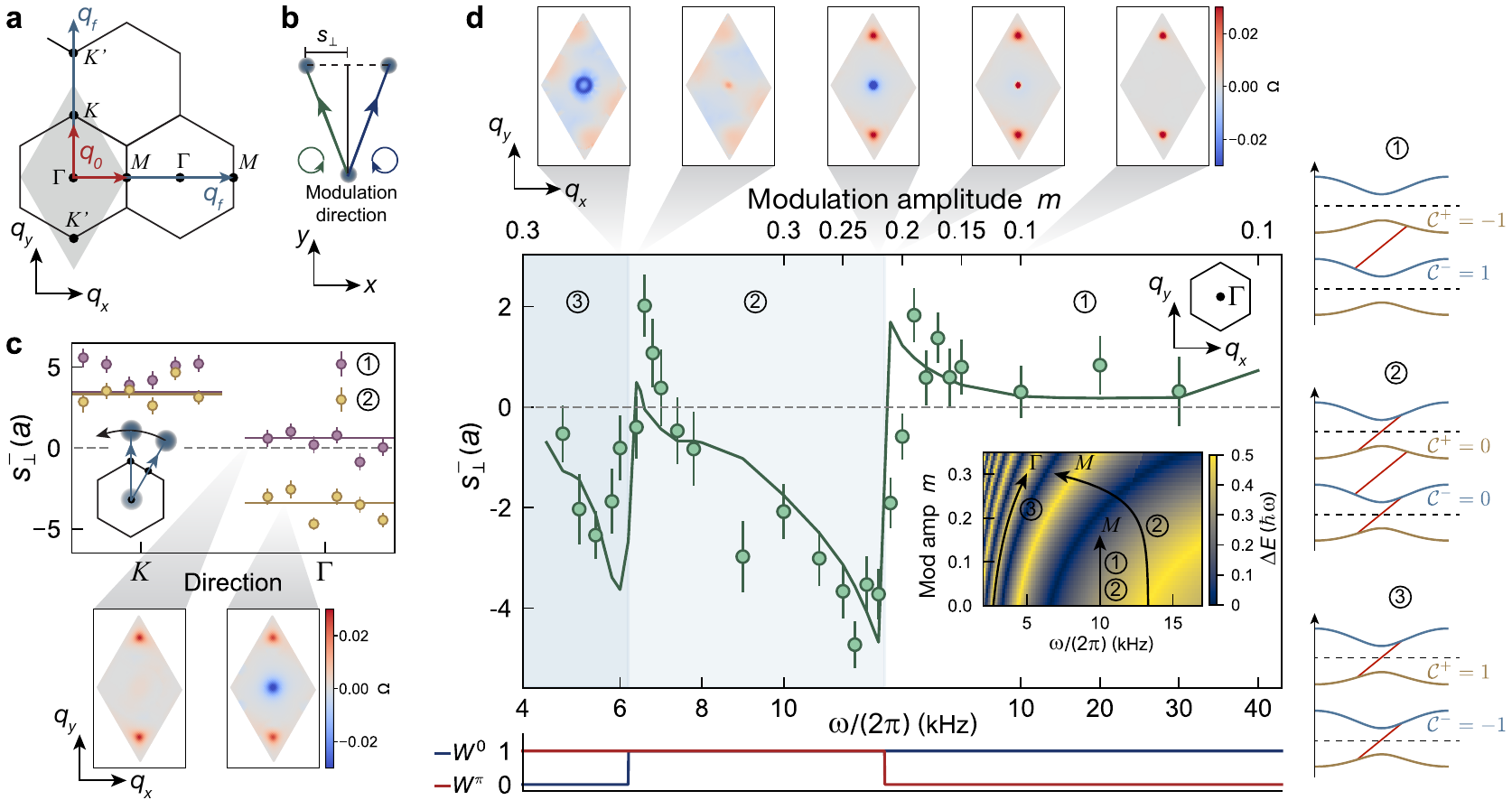}
\vspace{-0.cm} \caption{Schematics and experimental results for the local Hall deflections $s^-_\perp$ to probe the local Berry curvature distribution $\Omega$.
\textbf{a}. Schematics of the paths traversed in quasimomentum space for measuring the transverse deflections along the $K$- and $\Gamma$-direction ($q_0\rightarrow q_f$) and the first BZ (grey shaded area).
\textbf{b}. Definition of the differential transverse deflection $s^\mu_{\perp}$: The CoM positions for left- and right circular modulation as well as the starting position are measured as the mean values over 30-40 individual experimental realizations for each point. The bisecting line of the opening angle formed by the two paths defines an axis and the differential deflection is the distance of each final position to this axis (being the same for both modulation directions per definition).
\textbf{c}. Measured transverse deflections $s^-_{\perp}$ along all $K$- and $\Gamma$-directions (inset) with $F a/h = 204\,\text{Hz}$ for: $\omega/2 \pi = 16\,\text{kHz}$, $m=0.25$ (regime \ding{172}) and $\omega/2 \pi = 10\,\text{kHz}$, $m=0.24$ (regime \ding{173}). 
The solid lines are theoretical calculations (Supplementary Information) and the errorbars denote the standard error of the mean (SEM).
Lower panels: Calculated Berry curvature for the same modulation parameters in the first BZ.  
\textbf{d}. Main panel: Measured transverse deflections along the $\Gamma$-direction along the path in Fig.~\ref{Fig_2}a with $F a/h = 170\,\text{Hz}$, for $q_0\rightarrow q_{\text{eff}} \approx 1.25 \sqrt{3}\,k_L$ (regime \ding{172} and \ding{173}) and for $\Gamma \rightarrow q_{\text{eff}} \approx 0.93 \sqrt{3} \, k_L$ (regime \ding{174}). 
Errorbars indicate the SEM. Upper panels: Calculated Berry curvature distributions in the first BZ.
Inset: Change of modulation amplitude and frequency during the ramp-up in the three different regimes (denoted by the numbers) while changing the quasimomentum from $\Gamma$ to the point denoted at the end of the ramp-up path (Methods), the parameters are $\omega/(2 \pi) = \{5, 7, 10\}\,\text{kHz}$, $m=\{0.3, 0.3, 0.16\}$.  
Lower panel: Winding numbers deduced from the measured transverse deflections.
Right panel: Schematics of the energy bands in the different topological phases with the corresponding Chern numbers and edge modes (red lines).}
\label{Fig_3}
\end{figure*}


The change of topological invariants across the phase transition is determined by the signed topological charge $Q_s^j$ 
associated with the band touching singularity, which occurs at $(\mathbf{q}_s,\lambda_s)$, 
in the abstract 3D parameter space 
spanned by the quasimomentum $\mathbf{q}$ and the modulation parameter $\lambda$, which 
parametrizes the path through the phase diagram (white arrows in Fig.~\ref{Fig_2}a).
For a generic case of linear band touching points as in our model, 
the topological charge is $Q_s^j=\pm 1$~\cite{simon_holonomy_1983,bellissard_change_1995}. 
Other examples of parabolic band touchings have been found in 
the context of quantum chaos~\cite{leboeuf_topological_1992} and Harper-like models~\cite{barelli_semiclassical_1992}.

The value of the topological charge uniquely defines the change of the topological invariants 
\begin{equation}
W^j_{\lambda_s+\varepsilon} = W^j_{\lambda_s-\varepsilon} + Q_s^j
\label{eq:topcharge}
\end{equation}
across the phase transition, where $\varepsilon$ is a small parameter.
As a result, we can determine the winding numbers of any topological phase by tracking the number of gap-closing points
and characterizing the associated topological charge along a smooth path parametrized by $\lambda$.
Note that there can be several degenerate singularities $Q_s^j$ at the same parameters $(\mathbf{q}_s,\lambda_s)$,
which results in $|\Delta W_{\lambda_s}^j|=|W^j_{\lambda_s+\varepsilon}-W^j_{\lambda_s-\varepsilon}|>1$. 
Such a situation could be always identified experimentally by applying a small perturbation, which lifts the degeneracy and results in
isolated phase transitions~\cite{bellissard_change_1995,nathan_topological_2015}.

The high-frequency limit $\omega \rightarrow \infty$ can always be mapped
onto a static Hamiltonian via the rotating-wave approximation, in which case the winding number $W^\pi$ between Floquet zones is necessarily trivial, $W^\pi \xrightarrow[]{\omega \rightarrow \infty} 0$. 
In our model this limit corresponds to the Haldane phase with $\mathcal{C}^{\pm}=\mp1$ that has 
has been studied extensively both in theory and experiment~\cite{oka_photovoltaic_2009,rechtsman_photonic_2013,jotzu_experimental_2014}.
The set of topological invariants characterizing the high-frequency limit 
of our time-dependent model is thus, $(W^0,W^{\pi})=(1,0)$ and we can
characterize the other topological phases by 
tracking the evolution of $W^j$ along the path shown in Fig.~\ref{Fig_2}a by extracting the topological charges $Q_s^j$ and using Eq.~(\ref{eq:topcharge}). 
As derived in the Supplementary Information, the value of the topological charge is determined by the sign of the local Berry curvature $\Omega^{\pm}(\mathbf{q}_s)$, concentrated at the band touching singularity. 
The concentrated Berry curvature is associated with a $\pi$-Berry flux in quasimomentum space near the gap-closing singularity and the sign of the topological charge is given by the change of sign of this $\pi$-flux across the phase transition (upper panels in Fig.~\ref{Fig_3}d).
In the experiment we investigate the local Berry curvature distribution via Hall drift measurements~\cite{jotzu_experimental_2014,aidelsburger_measuring_2015} close to the singularity. 
Due to the finite width of the momentum distribution of the BEC, we obtain a signal that is averaged and weighted according to the momentum-space profile (Supplementary Information). 
Nonetheless, the change in sign of the Berry flux across the phase transition can be unambiguously defined from a change in sign of the 
measured deflections $s^\mu_\perp$ across the phase transition 

\begin{align}
\label{eq:defTC}
&Q^0_s = \text{sgn}\left( \Delta s^-_\perp(\mathbf{q}_s )\right), \quad
Q^\pi_s = -\text{sgn}\left( \Delta s^-_\perp(\mathbf{q}_s )\right) , \\
&\Delta s^-_\perp(\mathbf{q}_s) =  \text{sgn}[s^-_\perp(\mathbf{q}_s,\lambda_s+\varepsilon)]- \text{sgn}[s^-_\perp(\mathbf{q}_s,\lambda_s-\varepsilon)], \notag
\end{align}

\noindent where the transverse deflection $s^\mu_\perp$ is defined with respect to the energy band the atoms are prepared in.


The sequence starts by applying a force to the atoms by linear acceleration of the lattice (Methods), 
to adiabatically move the wavepacket in quasimomentum space. 
The Berry curvature acts as an effective magnetic field in quasimomentum-space, adding a transverse component to the atom's velocity~\cite{price_mapping_2012,dauphin_extracting_2013,jotzu_experimental_2014,aidelsburger_measuring_2015}. 
This anomalous velocity is directly proportional to the Berry curvature and gives rise to a deflection perpendicular to the direction of the force (Eq.~(S.3) in the Supplementary Information). We typically move the atoms along high-symmetry paths: $\Gamma - M - \Gamma$ ($\Gamma$-direction) and $\Gamma - K - K'$ ($K$-direction) between $q_0$ and $q_f$ (Fig.~\ref{Fig_3}a), probing regions with non-vanishing Berry curvature at the $\Gamma$- and $K$-points.

Since the force is generated by lattice acceleration, there is an additional longitudinal velocity component, which gives rise to large displacements in the direction of the force $F$.
This leads to a restoring force from the harmonic trap, which reduces the final longitudinal quasimomentum to $q_{\text{eff}}<q_f$, whereas the effect along the transverse direction can be neglected (Supplementary Information).
We record the final position of the cloud after applying the force by taking insitu absorption images and determine the center-of-mass (CoM) position by fitting a two-dimensional (2D) Gaussian function. 
We perform the same measurement for opposite chirality of the modulation to evaluate the differential deflection $s^\mu_{\perp}$ (Fig.~\ref{Fig_3}b), which is more robust to systematic deviations in the CoM position of the cloud.


To characterize the Haldane regime and quantitatively validate our experimental approach, we probed the Berry curvature almost in the entire first BZ by measuring deflections along all $K$- and $\Gamma$-directions in the range $q_0=0.5 \sqrt{3} \, k_L \rightarrow q_{\text{eff}} \approx 1.33 \sqrt{3} \, k_L$ (Fig.~\ref{Fig_3}c).
The Berry curvature around $\Gamma$ as well as around both Dirac points is traversed once by all atoms.
We find a good agreement between our experiment and ab initio numerical simulations taking into account the finite Gaussian width $\sigma$ of the momentum distribution (Supplementary Information), which was independently calibrated for each data set and lies in the range $\sigma \in [0.137, 0.168] \, k_L$.
In particular, we observe positive deflections throughout the whole BZ, which coincide well with theoretical calculations that employ a band with $\mathcal{C}^- = 1$.
In contrast, we find distinct negative deflections along all $\Gamma$-directions in the anomalous regime, which is consistent with a quasienergy band with trivial Chern number $\mathcal{C}^- = 0$.


In Fig.~\ref{Fig_3}d we show a complete scan of the transverse deflections at the $\Gamma$-point, where the gap closings occur, across the three different regimes. We traverse the Berry curvature around the $\Gamma$-point approximately once with the full cloud. The start and end point of the path in reciprocal space as well as the ramp-up of the modulation depend on the parameter regime (inset of Fig.~\ref{Fig_3}d and Methods). 
At larger modulation frequencies, we measure a slight positive deflection, as expected in the Haldane regime, which then grows when approaching the gap closing point. At the phase transition it suddenly changes to negative values, indicating a change of the winding number by $ \text{sgn}\left( \Delta s^-_\perp(\Gamma)\right)=-1$. Combined with the previous energy gap measurements, which indicated a gap closing point in the $\pi$-gap, we conclude that $Q_s^\pi=+1$, according to Eq.~(\ref{eq:defTC}), and that the winding number $W^\pi$ changes from $0\rightarrow 1$ (Fig.~\ref{Fig_2}c bottom) across the first phase transition \ding{172}-\ding{173}. This signals the transition to the anomalous phase, where $\mathcal{C}^{\pm}=0$ and $(W^0,W^\pi)=(1,1)$. 

Approaching the second phase transition \ding{173}-\ding{174}, the deflection $s^-_{\perp}$ starts to decrease due to the spreading of the negative Berry curvature in quasimomentum space. Shortly before the gap-closing point at $g^0$, the deflection turns positive and jumps to negative values after the transition. According to Eq.~(\ref{eq:defTC}) the topological charge at this transition is $Q_s^0=-1$, signaling a change of $W^{0}$ from $1\rightarrow 0$. At this transition we enter the third topological phase with $\mathcal{C}^{\pm}=\pm 1$ and  $(W^0,W^\pi)=(0,1)$, which is characterized by topological quasienergy bands similar to the Haldane phase with the difference, that there is an edge mode located between FBZs. 
Whereas in general the measured transverse deflections coincide well with the theoretical calculations, deviations can be observed near the phase transitions, where the quasienergy gap $\Delta E$ vanishes. We attribute these to non-adiabatic excitations to higher bands, which become important if $Fa\gtrsim \Delta E$.

\begin{figure}[t]
\includegraphics{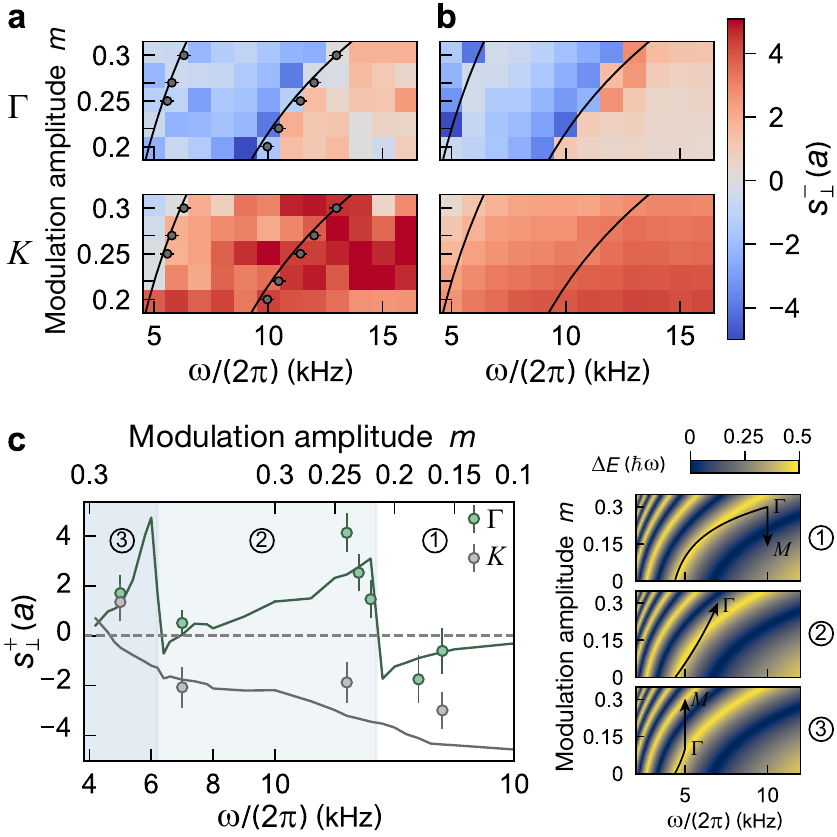}
\vspace{-0.cm} \caption{Transverse deflections in both bands in the three different regimes.
\textbf{a}. Measured transverse deflections $s_{\perp}^-$ at the $\Gamma$- (upper panel) and $K$-point (lower panel); the strength of the force was adapted to the different bandgaps, also changing the effective final quasimomentum (Methods). 
The SEM is on average $\bar{\Sigma}_{\Gamma}=0.72 (4)\, a$ and $\bar{\Sigma}_{K}=0.78 (5) \, a$. 
The dark gray data points correspond to the phase transition points obtained from linear fits to bandgap measurements at $\Gamma$ where the errorbars represent the stepsize of the modulation frequency used in the St\"uckelberg interferometry (Supplementary Information). 
The solid lines show the theoretical phase boundaries. 
\textbf{b}. Calculated transverse deflections, including effects due to the harmonic trap and the calibrated momentum-space widths. Solid lines are the same as in \textbf{a}. 
\textbf{c}. Left panel: Transverse deflections $s_{\perp}^+$ in the second band for $q_0\rightarrow q_{\text{eff}} \approx 1.25 \sqrt{3} \, k_L$ and for $\Gamma\rightarrow q_{\text{eff}} \approx 0.93 \, k_L$ in regime \ding{173} along $\Gamma$; $F a/h =  170\,\text{Hz}$. 
The modulation parameters are varied along the path depicted in Fig.~\ref{Fig_2}a, the shadings and numbers indicate different topological phases. Errorbars indicate the SEM.
Right panel: Variation of the modulation frequency and amplitude and final quasimomentum during the ramp-up and loading into the second band in the different regimes, identified by the numbers on the right; $\omega/(2 \pi) = \{5, 7, 10\} \,\text{kHz}$, $m=\{0.3, 0.3, 0.16\}$ (top to bottom).}
\label{Fig_4}
\end{figure}

To confirm our observations, we further probed the energy gap closings as well as the local Berry curvature for a larger range of modulation parameters (Fig.~\ref{Fig_4}a). 
Over the full parameter regime shown here and in Fig.~\ref{Fig_3}d, the measured deflections are in quantitative agreement with numerical simulations (Fig.~\ref{Fig_4}b) taking into account the finite momentum-space width and the harmonic trap, without free parameters. 
We further show the experimentally determined phase boundaries, which were extracted from energy gap measurements at the $\Gamma$-point (Supplementary Information).


The periodic nature of the Floquet bands further allows for direct loading of the atoms into the second band, when the modulation parameters lie in the anomalous regime during the complete ramp-up of the modulation (right panel of Fig.~\ref{Fig_4}c and Methods). This can be understood as follows:
In the limit of $m \rightarrow 0$ the quasienergy bands are multiple copies of the static energy bands separated  by $\hbar \omega$. Increasing the modulation amplitude opens gaps at the crossing points and hybridizes the two lowest bands. 
A condensate located at the $\Gamma$-point initially and during the ramp-up of the modulation amplitude, is now adiabatically transferred in the upper Floquet energy band, if the modulation frequency $\omega$ is kept within $[3.3, 6.6] \,\text{kHz}$. 
To also probe the other two regimes we divided the ramp-up into two parts, first loading the second band in the anomalous regime as described above, but with a smaller or larger amplitude, and then ramping to the final amplitude while moving by $q_0$ to avoid the gap closing points. 
Using these different ramp-up schemes we were able to probe the Berry curvature of the second band in the different regimes around the $\Gamma$- and $K$-points, plotted in the left panel of Fig.~\ref{Fig_4}c. 
We indeed observe an inversion of the Berry curvature, as expected from theory.


In this work we have presented the first experimental realization of anomalous Floquet systems with cold atoms and its complete characterization using bulk winding numbers $W$, which were deduced by combining energy gap measurements with local Hall deflections. 
The experimental data is in excellent agreement with numerical band structure calculations involving the first six bands of the modulated lattice over a wide range of parameters explicitly probing three distinct topological regimes. 
Moreover, the large degree of control of our experimental setup and observables facilitated an independent probe of the geometric properties of the first excited band. 
The successful realization of different topological systems with rather long lifetimes in the non-interacting limit, opens the door to a variety of interesting phenomena, for instance, the properties of edge modes at the boundaries~\cite{buchhold_effects_2012,goldman_direct_2013,reichl_floquet_2014} or the interplay of disorder and topology in driven systems~\cite{titum_disorder-induced_2015,titum_anomalous_2016}, to name only a few. 
The many-body regime provides a particularly rich experimental and theoretical playground~\cite{rudner_floquet_2019}.
Intriguingly, it has been shown that an anomalous Floquet insulator may exhibit a remarkable robustness against disorder~\cite{nathan_anomalous_2019}, which holds great promise for realizing quantized transport at high temperatures. 
This is due to the non-zero winding number of the Floquet spectrum, which cannot be annihilated even if all bulk states are localized, 
in contrast to conventional quantum Hall insulators. 
Moreover, in the anomalous phase one of the quasienergy bands exhibits a moatlike dispersion, i.e. a ring-shaped minimum of near-degenerate states, which can give rise to exotic many-body phenomena~\cite{gopalakrishnan_universal_2011,sedrakyan_composite_2012,sedrakyan_statistical_2015}.


\paragraph*{\textbf{Acknowledgements}}
We thank Jean Bellissard, Erez Berg, Jean Dalibard, Eugene Demler and Netanel Lindner for inspiring discussions. The research in Munich and Dresden was funded by the Deutsche Forschungsgemeinschaft (DFG, German Research Foundation) via Research Unit FOR 2414 under project number 277974659. The work in Munich was further supported under Germany's Excellence Strategy -- EXC-2111 -- 39081486. F.N.\"{U}. further acknowledges support from EPSRC Grant No. EP/P009565/1. The work in Belgium was supported by the ERC Starting Grant TopoCold, and the Fonds De La Recherche Scientifique (FRS-FNRS, Belgium).

\paragraph*{\textbf{Author contributions}}
K.W., C.B., M.D.L., N.G. and M.A. designed the modulation scheme and benchmarked the deflection measurement. F.N.\"U., A.E. and M.A. devised the protocol to extract the set of topological invariants. K.W. and C.B. performed the experiment, analyzed the data and performed the numerical simulations with M.A. All authors contributed to the writing of the manuscript and the discussion of the results.

\paragraph*{\textbf{Data availability}}
The data that support the plots within this paper and other findings of this study are available from the corresponding author upon reasonable request.

\paragraph*{\textbf{Code availability}}
The code that supports the plots within this paper are available from the corresponding author upon reasonable request.

\paragraph*{\textbf{Competing interests}}
The authors declare no competing interests.


\section*{Methods}


\paragraph*{St\"uckelberg interferometry:} The energy gaps between the two lowest bands are measured using St\"uckelberg interferometry~\cite{zenesini_observation_2010,kling_atomic_2010}. 
We start by loading the atoms into the lowest band of the static lattice at $\Gamma$ and then change their quasimomentum non-adiabatically to the point where the bandgap should be probed, which leads to coherent population of the second band.
The change in quasimomentum was carried out via linear frequency sweeps of two laser beams, enabling large forces in arbitrary directions in the 2D quasimomentum space. 
The modulation amplitude was ramped up linearly at the desired modulation frequency within five modulation cycles $T$ and the acceleration started meanwhile, such that the final quasimomentum was reached at the end of the ramp-up.
Then, the atoms were held at the final quasimomentum for integer multiples of the driving cycle and subsequently accelerated back to $\Gamma$ in the first BZ using the same force as before while ramping down the modulation inversely to the ramp-up. 
During the hold time, the atoms acquire a dynamical phase depending on the energy band and quasimomentum they are occupying. 
Driving back non-adiabatically recombines the populations in the two bands leading to oscillations of the band populations in time with a frequency given by the bandgap at the probed quasimomentum.

Considering the first order Floquet copies of the $s$-bands, there are two energy gaps that can be probed, the gap at zero energy, $g^0 (\mathbf{q})$, and between Floquet zones, $g^{\pi}(\mathbf{q})$, with $g^0(\mathbf{q})+g^{\pi}(\mathbf{q})=\hbar \omega$ at each point in quasimomentum space. 
Since the hold times are always integer multiples of the modulation cycle, the maximum gap frequency that can be measured with St\"uckelberg interferometry is $\omega_{\text{max}}=\omega/2$, which would correspond to sampling the cosine-wave with two points per oscillation. This means that at every $\mathbf{q}$ we always measure the smaller gap out of $g^0(\mathbf{q})$ and $g^{\pi}(\mathbf{q})$, enabling us to differentiate between them as described in the main text.

At certain modulation parameters in the anomalous regime excitations to the second band can occur during the ramp-up due to energy gap closings. This leads only to an offset phase in the St\"uckelberg oscillations and does not change their frequency. 
The relative population in the lowest band $n^1$ is measured by taking absorption images after performing bandmapping and a time-of-flight (TOF) of $3.5 \, \text{ms}$ (Supplementary Information). 
We average the relative population over 3-4 independent experimental realizations for each hold time and extract the oscillation frequency from a fit. 
Due to the periodic nature of the quasienergy bands, there might be excitations to Floquet copies of the $p$-bands, being suppressed with the corresponding Floquet order. 
This would lead to an oscillation with multiple frequencies which we take into account by fitting a sum of two cosine functions. 
We also include a possible damping of the oscillation due to dynamical instabilities and atom loss:

\begin{align}
n^1(t) &= \text{e}^{-(t-t_1)\gamma}A_1\text{cos}(\omega_1 (t-t_1)) \notag \\ 
&+A_2\text{cos}(\omega_2 (t-t_2)) + n_0,
\label{Eq_n1}
\end{align}

\noindent where $A_1$, $A_2$, $\omega_1$, $\omega_2$, $t_1$, $t_2$ and $\gamma$ are free fit parameters and the main oscillation frequency is defined as having the larger relative amplitude. 
For most parameters, the contribution from the second frequency is negligible, it mainly plays a role in the anomalous regime close to the phase transition. 
In Fig.~S4 an example of population oscillations at $\Gamma$ is shown together with its Fast Fourier transform (FFT), which clearly shows the closing and opening of the energy gaps.  


\paragraph*{Deflection measurements:} For the deflection measurements the force is also applied using lattice acceleration but now it is small compared to the energy gaps, in order to adiabatically move inside a single band. 
Here, the ramp-up of the modulation depends on the band and topological regime that is probed. 
When measuring in the first band at modulation frequencies $\omega/(2 \pi) \geq 8\,\text{kHz}$, being in the Haldane and anomalous regime, we ramp up the modulation amplitude at the desired frequency while moving the condensate to $q_0 = 0.5 \sqrt{3} \, k_L$ (the $M$-point when driving along the $\Gamma$-direction). At this point the final modulation parameters are reached and we continue to accelerate to the final quasimomentum $q_f$ for the local deflection measurement. 
The ramp-up time is chosen as the time needed to change the quasimomentum of the cloud from $\Gamma$ to $q_0$, with the given force rounded up to full cycles of the modulation. We have verified independently that the magnitude of the force leads to negligible excitations (Fig.~S2b).
By moving away from the $\Gamma$-point we avoid the gap closing point, which would result in excitations to the second band. 
We verified that the points in quasimomentum space where the bands have crossed and hybridized (the ring-shaped minimum in the anomalous phase) are always located away from the outer edge of the moving cloud defined by its Gaussian width in reciprocal space, meaning that during the ramp-up the cloud does not traverse regions with non-zero Berry curvature. 
Hence, we effectively probe the Berry curvature along paths in quasimomentum space starting at a distance of $q_0$ from the $\Gamma$-point, which was also used in the calculations.
For smaller modulation frequencies the phase transition takes place at earlier times during the ramp-up and the lowest frequency for which we used this scheme was set to $8\,\text{kHz}$ to ensure that the energy gaps the atoms see during the ramp-up are always larger than $\Delta E/h \approx 500\,\text{Hz}$. 
From the good agreement between the measured deflections and the theoretical calculations we can confirm that for most modulation parameters there can only be minor excitations to higher bands and the Berry curvature probed during the ramp-up is negligible. 
We also verified the latter experimentally by measuring the transverse deflection when accelerating by $q_0$ along the $\Gamma$- and $K$-directions (see Fig.~S2a).

For measurements in the anomalous regime with $\omega/(2 \pi) < 8\,\text{kHz}$ we ramped the modulation amplitude and frequency simultaneously, starting at $f_0 = 13.3\,\text{kHz}$ which is the modulation frequency at which the relative gap in the Haldane phase is maximal for $m \rightarrow 0$. 
The modulation amplitude was increased linearly in time, while again accelerating the atoms to up to $q_0$, and the frequency was changed exponentially (see also inset of Fig.~\ref{Fig_3}d): 
\begin{align}
\omega(t)/(2 \pi) = \frac{f_f-f_0}{e^{m_f \cdot p} - 1} \left(e^{\frac{m_f}{t_r} t \cdot p} - 1\right)+f_0,
\label{Eq_rampup}
\end{align}

\noindent where the final frequency and amplitude are denoted by $f_f$ and $m_f$, respectively, $t_r$ is the ramp-up time and $p=20$ for the ramp-up in the anomalous regime. 
The functional form was motivated by the shape of the phase transition lines, which approximately follow $f(m) \propto e^{m}$. 
When using this chirped ramp-up, the time was chosen as being close to the traversing time to $q_0$ but ending at full cycles of the modulation.
We verified numerically that for all modulation parameters probed here, the minimal gap and Berry curvature during the ramp-up fulfill the requirements stated above.

In the third regime we start at $f_0=2.7\,\text{kHz}$, which is again at the maximal relative gap, and ramp up according to Eq.~(\ref{Eq_rampup}), but now the parameter $p$ is fitted to maximize the (absolute) energy gap during the ramp-up and the atoms are held at $\Gamma$. 
When probing the Berry curvature around $\Gamma$, we moved by $q_{\text{eff}} \approx 0.93 \sqrt{3} \, k_L$, so along $\Gamma - M - \Gamma$, which is equivalent to $M - \Gamma - M$ for the modulation parameters used here, as verified numerically. 
This is not true along the $K$-directions, where we determined the final deflection from $q_0$ to $q_f$ by measuring along the whole path and subtracting the measured deflection when only moving up to $q_0$. 
Here, the ramp-up time was chosen to be similar to the traversing times used in the other regimes ($\approx 1.96 \, \text{ms}$) and ending with a full modulation cycle again.

To probe the second band, we always started in the anomalous regime, where it is connected to the first band of the static lattice for $m\rightarrow 0$ (see main text and right panels of Fig.~\ref{Fig_4}c). 
To probe the anomalous regime, the same ramp-up scheme was used as for the first band in the third regime, but now starting at $f_0=4.4\,\text{kHz}$. 
To reach the Haldane regime, we first performed a ramp in the anomalous regime from $f_0 = 4.4\,\text{kHz}$ to $m=0.3$ and $f_f$ at $\Gamma$ with $p$ being fitted to maximize the gap, and then decreased the amplitude linearly to $m_f$, while moving to $q_0$, avoiding the gap closing points. 
In this case, the atoms do traverse Berry curvature during the second part of the ramp-up, so we determined again the final deflection by subtracting the deflection measured up to $q_0$, now along all directions. 
In the third regime we used a similar procedure, now ramping up to $m=0.1$ at $\Gamma$ and then linearly increasing the amplitude to $m_f$.

For the frequency scans presented in Fig.~\ref{Fig_4}a, the force was adapted to the energy gaps: 
For $m=0.25$ and $\omega/(2 \pi) \geq 8\,\text{kHz}$ we used $Fa/h = 272\,\text{Hz}$ probing $q_0 \rightarrow q_{\text{eff}} \approx 1.40 \sqrt{3} \, k_L$; for $m=0.3$ in the same frequency range, $Fa/h = 341\,\text{Hz}$ and $q_0 \rightarrow q_{\text{eff}} \approx 1.33 \sqrt{3} \, k_L$. In regime \ding{174} we probed $\Gamma \rightarrow q_{\text{eff}} \approx 0.95 \sqrt{3}\,k_L$ along the $\Gamma$-direction using $Fa/h = 204\,\text{Hz}$. In all other cases, $F a /h = 204\,\text{Hz}$ and $q_0 \rightarrow q_{\text{eff}} \approx 1.33 \sqrt{3}\, k_L$.

For all modulation parameters (including the parameter scans in Fig.~\ref{Fig_3}d and Fig.~\ref{Fig_4}c) except for $m=0.3$ and $\omega/(2 \pi) \geq 8\,\text{kHz}$ and when probing the first band in regime \ding{173} or the second band in regime \ding{174} along $\Gamma$, the final quasimomentum was set to $q_f= 1.5 \sqrt{3}\,k_L$ by the lattice acceleration, but the effective values $q_{\text{eff}}$ are reduced due to the restoring force of the harmonic trap (Supplementary Information), which increases for larger longitudinal displacements and hence smaller forces. 
In the case of $m=0.3$ and $\omega/(2 \pi) \geq 8\,\text{kHz}$ this was taken into account in the experiment and $q_f = 1.38 \sqrt{3} \,k_L$ was programmed yielding $q_{\text{eff}} \approx 1.33 \sqrt{3} \, k_L$ similar to the other measurements. 
The effective distance traversed during the ramp-up remains $q_0 \approx 0.5 \sqrt{3} \, k_L$, since the real space displacement is still small here. 
For the theoretical calculations, the full equations of motion were solved numerically, including the harmonic trap and the band dispersion as well as the measured momentum space widths in all cases (Supplementary Information). 


\begin{thebibliography}{55}%
\makeatletter
\providecommand \@ifxundefined [1]{%
 \@ifx{#1\undefined}
}%
\providecommand \@ifnum [1]{%
 \ifnum #1\expandafter \@firstoftwo
 \else \expandafter \@secondoftwo
 \fi
}%
\providecommand \@ifx [1]{%
 \ifx #1\expandafter \@firstoftwo
 \else \expandafter \@secondoftwo
 \fi
}%
\providecommand \natexlab [1]{#1}%
\providecommand \enquote  [1]{``#1''}%
\providecommand \bibnamefont  [1]{#1}%
\providecommand \bibfnamefont [1]{#1}%
\providecommand \citenamefont [1]{#1}%
\providecommand \href@noop [0]{\@secondoftwo}%
\providecommand \href [0]{\begingroup \@sanitize@url \@href}%
\providecommand \@href[1]{\@@startlink{#1}\@@href}%
\providecommand \@@href[1]{\endgroup#1\@@endlink}%
\providecommand \@sanitize@url [0]{\catcode `\\12\catcode `\$12\catcode
  `\&12\catcode `\#12\catcode `\^12\catcode `\_12\catcode `\%12\relax}%
\providecommand \@@startlink[1]{}%
\providecommand \@@endlink[0]{}%
\providecommand \url  [0]{\begingroup\@sanitize@url \@url }%
\providecommand \@url [1]{\endgroup\@href {#1}{\urlprefix }}%
\providecommand \urlprefix  [0]{URL }%
\providecommand \Eprint [0]{\href }%
\providecommand \doibase [0]{http://dx.doi.org/}%
\providecommand \selectlanguage [0]{\@gobble}%
\providecommand \bibinfo  [0]{\@secondoftwo}%
\providecommand \bibfield  [0]{\@secondoftwo}%
\providecommand \translation [1]{[#1]}%
\providecommand \BibitemOpen [0]{}%
\providecommand \bibitemStop [0]{}%
\providecommand \bibitemNoStop [0]{.\EOS\space}%
\providecommand \EOS [0]{\spacefactor3000\relax}%
\providecommand \BibitemShut  [1]{\csname bibitem#1\endcsname}%
\let\auto@bib@innerbib\@empty
\bibitem [{\citenamefont {Goldman}\ and\ \citenamefont
  {Dalibard}(2014)}]{goldman_periodically_2014}%
  \BibitemOpen
  \bibfield  {author} {\bibinfo {author} {\bibfnamefont {N.}~\bibnamefont
  {Goldman}}\ and\ \bibinfo {author} {\bibfnamefont {J.}~\bibnamefont
  {Dalibard}},\ }\bibfield  {title} {\enquote {\bibinfo {title} {Periodically
  {Driven} {Quantum} {Systems}: {Effective} {Hamiltonians} and {Engineered}
  {Gauge} {Fields}},}\ }\href {\doibase 10.1103/PhysRevX.4.031027} {\bibfield
  {journal} {\bibinfo  {journal} {Phys. Rev. X}\ }\textbf {\bibinfo {volume}
  {4}},\ \bibinfo {pages} {031027} (\bibinfo {year} {2014})}\BibitemShut
  {NoStop}%
\bibitem [{\citenamefont {Bukov}\ \emph {et~al.}(2015)\citenamefont {Bukov},
  \citenamefont {D'Alessio},\ and\ \citenamefont
  {Polkovnikov}}]{bukov_universal_2015}%
  \BibitemOpen
  \bibfield  {author} {\bibinfo {author} {\bibfnamefont {M.}~\bibnamefont
  {Bukov}}, \bibinfo {author} {\bibfnamefont {L.}~\bibnamefont {D'Alessio}}, \
  and\ \bibinfo {author} {\bibfnamefont {A.}~\bibnamefont {Polkovnikov}},\
  }\bibfield  {title} {\enquote {\bibinfo {title} {Universal high-frequency
  behavior of periodically driven systems: from dynamical stabilization to
  {Floquet} engineering},}\ }\href {\doibase 10.1080/00018732.2015.1055918}
  {\bibfield  {journal} {\bibinfo  {journal} {Adv. Phys.}\ }\textbf {\bibinfo
  {volume} {64}},\ \bibinfo {pages} {139--226} (\bibinfo {year}
  {2015})}\BibitemShut {NoStop}%
\bibitem [{\citenamefont {Eckardt}(2017)}]{eckardt_colloquium_2017}%
  \BibitemOpen
  \bibfield  {author} {\bibinfo {author} {\bibfnamefont {A.}~\bibnamefont
  {Eckardt}},\ }\bibfield  {title} {\enquote {\bibinfo {title} {Colloquium:
  {Atomic} quantum gases in periodically driven optical lattices},}\ }\href
  {\doibase 10.1103/RevModPhys.89.011004} {\bibfield  {journal} {\bibinfo
  {journal} {Rev. Mod. Phys.}\ }\textbf {\bibinfo {volume} {89}},\ \bibinfo
  {pages} {011004} (\bibinfo {year} {2017})}\BibitemShut {NoStop}%
\bibitem [{\citenamefont {Struck}\ \emph {et~al.}(2011)\citenamefont {Struck},
  \citenamefont {\"Olschl\"ager}, \citenamefont {Le~Targat}, \citenamefont
  {Soltan-Panahi}, \citenamefont {Eckardt}, \citenamefont {Lewenstein},
  \citenamefont {Windpassinger},\ and\ \citenamefont
  {Sengstock}}]{struck_quantum_2011}%
  \BibitemOpen
  \bibfield  {author} {\bibinfo {author} {\bibfnamefont {J.}~\bibnamefont
  {Struck}}, \bibinfo {author} {\bibfnamefont {C.}~\bibnamefont
  {\"Olschl\"ager}}, \bibinfo {author} {\bibfnamefont {R.}~\bibnamefont
  {Le~Targat}}, \bibinfo {author} {\bibfnamefont {P.}~\bibnamefont
  {Soltan-Panahi}}, \bibinfo {author} {\bibfnamefont {A.}~\bibnamefont
  {Eckardt}}, \bibinfo {author} {\bibfnamefont {M.}~\bibnamefont {Lewenstein}},
  \bibinfo {author} {\bibfnamefont {P.}~\bibnamefont {Windpassinger}}, \ and\
  \bibinfo {author} {\bibfnamefont {Klaus}\ \bibnamefont {Sengstock}},\
  }\bibfield  {title} {\enquote {\bibinfo {title} {Quantum {Simulation} of
  {Frustrated} {Classical} {Magnetism} in {Triangular} {Optical} {Lattices}},}\
  }\href {https://science.sciencemag.org/content/333/6045/996} {\bibfield
  {journal} {\bibinfo  {journal} {Science}\ }\textbf {\bibinfo {volume}
  {333}},\ \bibinfo {pages} {996--999} (\bibinfo {year} {2011})}\BibitemShut
  {NoStop}%
\bibitem [{\citenamefont {Aidelsburger}\ \emph {et~al.}(2011)\citenamefont
  {Aidelsburger}, \citenamefont {Atala}, \citenamefont {Nascimb\`ene},
  \citenamefont {Trotzky}, \citenamefont {Chen},\ and\ \citenamefont
  {Bloch}}]{aidelsburger_experimental_2011}%
  \BibitemOpen
  \bibfield  {author} {\bibinfo {author} {\bibfnamefont {M.}~\bibnamefont
  {Aidelsburger}}, \bibinfo {author} {\bibfnamefont {M.}~\bibnamefont {Atala}},
  \bibinfo {author} {\bibfnamefont {S.}~\bibnamefont {Nascimb\`ene}}, \bibinfo
  {author} {\bibfnamefont {S.}~\bibnamefont {Trotzky}}, \bibinfo {author}
  {\bibfnamefont {Y.-A.}\ \bibnamefont {Chen}}, \ and\ \bibinfo {author}
  {\bibfnamefont {I.}~\bibnamefont {Bloch}},\ }\bibfield  {title} {\enquote
  {\bibinfo {title} {Experimental {Realization} of {Strong} {Effective}
  {Magnetic} {Fields} in an {Optical} {Lattice}},}\ }\href {\doibase
  10.1103/PhysRevLett.107.255301} {\bibfield  {journal} {\bibinfo  {journal}
  {Phys. Rev. Lett.}\ }\textbf {\bibinfo {volume} {107}},\ \bibinfo {pages}
  {255301} (\bibinfo {year} {2011})}\BibitemShut {NoStop}%
\bibitem [{\citenamefont {Rechtsman}\ \emph {et~al.}(2013)\citenamefont
  {Rechtsman}, \citenamefont {Zeuner}, \citenamefont {Plotnik}, \citenamefont
  {Lumer}, \citenamefont {Podolsky}, \citenamefont {Dreisow}, \citenamefont
  {Nolte}, \citenamefont {Segev},\ and\ \citenamefont
  {Szameit}}]{rechtsman_photonic_2013}%
  \BibitemOpen
  \bibfield  {author} {\bibinfo {author} {\bibfnamefont {M.~C.}\ \bibnamefont
  {Rechtsman}}, \bibinfo {author} {\bibfnamefont {J.~M.}\ \bibnamefont
  {Zeuner}}, \bibinfo {author} {\bibfnamefont {Y.}~\bibnamefont {Plotnik}},
  \bibinfo {author} {\bibfnamefont {Y.}~\bibnamefont {Lumer}}, \bibinfo
  {author} {\bibfnamefont {D.}~\bibnamefont {Podolsky}}, \bibinfo {author}
  {\bibfnamefont {F.}~\bibnamefont {Dreisow}}, \bibinfo {author} {\bibfnamefont
  {S.}~\bibnamefont {Nolte}}, \bibinfo {author} {\bibfnamefont
  {M.}~\bibnamefont {Segev}}, \ and\ \bibinfo {author} {\bibfnamefont
  {A.}~\bibnamefont {Szameit}},\ }\bibfield  {title} {\enquote {\bibinfo
  {title} {Photonic {Floquet} topological insulators},}\ }\href {\doibase
  10.1038/nature12066} {\bibfield  {journal} {\bibinfo  {journal} {Nature}\
  }\textbf {\bibinfo {volume} {496}},\ \bibinfo {pages} {196--200} (\bibinfo
  {year} {2013})}\BibitemShut {NoStop}%
\bibitem [{\citenamefont {Hafezi}\ \emph {et~al.}(2013)\citenamefont {Hafezi},
  \citenamefont {Mittal}, \citenamefont {Fan}, \citenamefont {Migdall},\ and\
  \citenamefont {Taylor}}]{hafezi_imaging_2013}%
  \BibitemOpen
  \bibfield  {author} {\bibinfo {author} {\bibfnamefont {M.}~\bibnamefont
  {Hafezi}}, \bibinfo {author} {\bibfnamefont {S.}~\bibnamefont {Mittal}},
  \bibinfo {author} {\bibfnamefont {J.}~\bibnamefont {Fan}}, \bibinfo {author}
  {\bibfnamefont {A.}~\bibnamefont {Migdall}}, \ and\ \bibinfo {author}
  {\bibfnamefont {J.~M.}\ \bibnamefont {Taylor}},\ }\bibfield  {title}
  {\enquote {\bibinfo {title} {Imaging topological edge states in silicon
  photonics},}\ }\href {\doibase 10.1038/nphoton.2013.274} {\bibfield
  {journal} {\bibinfo  {journal} {Nature Photon.}\ }\textbf {\bibinfo {volume}
  {7}},\ \bibinfo {pages} {1001--1005} (\bibinfo {year} {2013})}\BibitemShut
  {NoStop}%
\bibitem [{\citenamefont {Roushan}\ \emph {et~al.}(2017)\citenamefont
  {Roushan}, \citenamefont {Neill}, \citenamefont {Megrant}, \citenamefont
  {Chen}, \citenamefont {Babbush}, \citenamefont {Barends}, \citenamefont
  {Campbell}, \citenamefont {Chen}, \citenamefont {Chiaro}, \citenamefont
  {Dunsworth}, \citenamefont {Fowler}, \citenamefont {Jeffrey}, \citenamefont
  {Kelly}, \citenamefont {Lucero}, \citenamefont {Mutus}, \citenamefont
  {O?Malley}, \citenamefont {Neeley}, \citenamefont {Quintana}, \citenamefont
  {Sank}, \citenamefont {Vainsencher}, \citenamefont {Wenner}, \citenamefont
  {White}, \citenamefont {Kapit}, \citenamefont {Neven},\ and\ \citenamefont
  {Martinis}}]{roushan_chiral_2017}%
  \BibitemOpen
  \bibfield  {author} {\bibinfo {author} {\bibfnamefont {P.}~\bibnamefont
  {Roushan}}, \bibinfo {author} {\bibfnamefont {C.}~\bibnamefont {Neill}},
  \bibinfo {author} {\bibfnamefont {A.}~\bibnamefont {Megrant}}, \bibinfo
  {author} {\bibfnamefont {Y.}~\bibnamefont {Chen}}, \bibinfo {author}
  {\bibfnamefont {R.}~\bibnamefont {Babbush}}, \bibinfo {author} {\bibfnamefont
  {R.}~\bibnamefont {Barends}}, \bibinfo {author} {\bibfnamefont
  {B.}~\bibnamefont {Campbell}}, \bibinfo {author} {\bibfnamefont
  {Z.}~\bibnamefont {Chen}}, \bibinfo {author} {\bibfnamefont {B.}~\bibnamefont
  {Chiaro}}, \bibinfo {author} {\bibfnamefont {A.}~\bibnamefont {Dunsworth}},
  \bibinfo {author} {\bibfnamefont {A.}~\bibnamefont {Fowler}}, \bibinfo
  {author} {\bibfnamefont {E.}~\bibnamefont {Jeffrey}}, \bibinfo {author}
  {\bibfnamefont {J.}~\bibnamefont {Kelly}}, \bibinfo {author} {\bibfnamefont
  {E.}~\bibnamefont {Lucero}}, \bibinfo {author} {\bibfnamefont
  {J.}~\bibnamefont {Mutus}}, \bibinfo {author} {\bibfnamefont {P.~J.~J.}\
  \bibnamefont {O'Malley}}, \bibinfo {author} {\bibfnamefont
  {M.}~\bibnamefont {Neeley}}, \bibinfo {author} {\bibfnamefont
  {C.}~\bibnamefont {Quintana}}, \bibinfo {author} {\bibfnamefont
  {D.}~\bibnamefont {Sank}}, \bibinfo {author} {\bibfnamefont {A.}~\bibnamefont
  {Vainsencher}}, \bibinfo {author} {\bibfnamefont {J.}~\bibnamefont {Wenner}},
  \bibinfo {author} {\bibfnamefont {T.}~\bibnamefont {White}}, \bibinfo
  {author} {\bibfnamefont {E.}~\bibnamefont {Kapit}}, \bibinfo {author}
  {\bibfnamefont {H.}~\bibnamefont {Neven}}, \ and\ \bibinfo {author}
  {\bibfnamefont {J.}~\bibnamefont {Martinis}},\ }\bibfield  {title} {\enquote
  {\bibinfo {title} {Chiral ground-state currents of interacting photons in a
  synthetic magnetic field},}\ }\href {\doibase 10.1038/nphys3930} {\bibfield
  {journal} {\bibinfo  {journal} {Nature Phys.}\ }\textbf {\bibinfo {volume}
  {13}},\ \bibinfo {pages} {146--151} (\bibinfo {year} {2017})}\BibitemShut
  {NoStop}%
\bibitem [{\citenamefont {McIver}\ \emph {et~al.}(2020)\citenamefont {McIver},
  \citenamefont {Schulte}, \citenamefont {Stein}, \citenamefont {Matsuyama},
  \citenamefont {Jotzu}, \citenamefont {Meier},\ and\ \citenamefont
  {Cavalleri}}]{mciver_light-induced_2020}%
  \BibitemOpen
  \bibfield  {author} {\bibinfo {author} {\bibfnamefont {J.~W.}\ \bibnamefont
  {McIver}}, \bibinfo {author} {\bibfnamefont {B.}~\bibnamefont {Schulte}},
  \bibinfo {author} {\bibfnamefont {F.-U.}\ \bibnamefont {Stein}}, \bibinfo
  {author} {\bibfnamefont {T.}~\bibnamefont {Matsuyama}}, \bibinfo {author}
  {\bibfnamefont {G.}~\bibnamefont {Jotzu}}, \bibinfo {author} {\bibfnamefont
  {G.}~\bibnamefont {Meier}}, \ and\ \bibinfo {author} {\bibfnamefont
  {A.}~\bibnamefont {Cavalleri}},\ }\bibfield  {title} {\enquote {\bibinfo
  {title} {Light-induced anomalous {Hall} effect in graphene},}\ }\href
  {https://www.nature.com/articles/s41567-019-0698-y} {\bibfield  {journal}
  {\bibinfo  {journal} {Nature Phys.}\ }\textbf {\bibinfo {volume} {16}},\
  \bibinfo {pages} {38--41} (\bibinfo {year} {2020})}\BibitemShut {NoStop}%
\bibitem [{\citenamefont {Aidelsburger}\ \emph {et~al.}(2018)\citenamefont
  {Aidelsburger}, \citenamefont {Nascimb\`ene},\ and\ \citenamefont
  {Goldman}}]{aidelsburger_artificial_2018}%
  \BibitemOpen
  \bibfield  {author} {\bibinfo {author} {\bibfnamefont {M.}~\bibnamefont
  {Aidelsburger}}, \bibinfo {author} {\bibfnamefont {S.}~\bibnamefont
  {Nascimb\`ene}}, \ and\ \bibinfo {author} {\bibfnamefont {N.}~\bibnamefont
  {Goldman}},\ }\bibfield  {title} {\enquote {\bibinfo {title} {Artificial
  gauge fields in materials and engineered systems},}\ }\href {\doibase
  10.1016/j.crhy.2018.03.002} {\bibfield  {journal} {\bibinfo  {journal} {C. R.
  Physique}\ }\textbf {\bibinfo {volume} {19}},\ \bibinfo {pages} {394--432}
  (\bibinfo {year} {2018})}\BibitemShut {NoStop}%
\bibitem [{\citenamefont {Cooper}\ \emph {et~al.}(2019)\citenamefont {Cooper},
  \citenamefont {Dalibard},\ and\ \citenamefont
  {Spielman}}]{cooper_topological_2019}%
  \BibitemOpen
  \bibfield  {author} {\bibinfo {author} {\bibfnamefont {N.~R.}\ \bibnamefont
  {Cooper}}, \bibinfo {author} {\bibfnamefont {J.}~\bibnamefont {Dalibard}}, \
  and\ \bibinfo {author} {\bibfnamefont {I.~B.}\ \bibnamefont {Spielman}},\
  }\bibfield  {title} {\enquote {\bibinfo {title} {Topological bands for
  ultracold atoms},}\ }\href {\doibase 10.1103/RevModPhys.91.015005} {\bibfield
   {journal} {\bibinfo  {journal} {Rev. Mod. Phys.}\ }\textbf {\bibinfo
  {volume} {91}},\ \bibinfo {pages} {015005} (\bibinfo {year}
  {2019})}\BibitemShut {NoStop}%
\bibitem [{\citenamefont {Aidelsburger}\ \emph {et~al.}(2013)\citenamefont
  {Aidelsburger}, \citenamefont {Atala}, \citenamefont {Lohse}, \citenamefont
  {Barreiro}, \citenamefont {Paredes},\ and\ \citenamefont
  {Bloch}}]{aidelsburger_realization_2013}%
  \BibitemOpen
  \bibfield  {author} {\bibinfo {author} {\bibfnamefont {M.}~\bibnamefont
  {Aidelsburger}}, \bibinfo {author} {\bibfnamefont {M.}~\bibnamefont {Atala}},
  \bibinfo {author} {\bibfnamefont {M.}~\bibnamefont {Lohse}}, \bibinfo
  {author} {\bibfnamefont {J.~T.}\ \bibnamefont {Barreiro}}, \bibinfo {author}
  {\bibfnamefont {B.}~\bibnamefont {Paredes}}, \ and\ \bibinfo {author}
  {\bibfnamefont {I.}~\bibnamefont {Bloch}},\ }\bibfield  {title} {\enquote
  {\bibinfo {title} {Realization of the {Hofstadter} {Hamiltonian} with
  {Ultracold} {Atoms} in {Optical} {Lattices}},}\ }\href {\doibase
  10.1103/PhysRevLett.111.185301} {\bibfield  {journal} {\bibinfo  {journal}
  {Phys. Rev. Lett.}\ }\textbf {\bibinfo {volume} {111}},\ \bibinfo {pages}
  {185301} (\bibinfo {year} {2013})}\BibitemShut {NoStop}%
\bibitem [{\citenamefont {Miyake}\ \emph {et~al.}(2013)\citenamefont {Miyake},
  \citenamefont {Siviloglou}, \citenamefont {Kennedy}, \citenamefont {Burton},\
  and\ \citenamefont {Ketterle}}]{miyake_realizing_2013}%
  \BibitemOpen
  \bibfield  {author} {\bibinfo {author} {\bibfnamefont {H.}~\bibnamefont
  {Miyake}}, \bibinfo {author} {\bibfnamefont {G.~A.}\ \bibnamefont
  {Siviloglou}}, \bibinfo {author} {\bibfnamefont {C.~J.}\ \bibnamefont
  {Kennedy}}, \bibinfo {author} {\bibfnamefont {W.~C.}\ \bibnamefont {Burton}},
  \ and\ \bibinfo {author} {\bibfnamefont {W.}~\bibnamefont {Ketterle}},\
  }\bibfield  {title} {\enquote {\bibinfo {title} {Realizing the {Harper}
  {Hamiltonian} with {Laser}-{Assisted} {Tunneling} in {Optical} {Lattices}},}\
  }\href {\doibase 10.1103/PhysRevLett.111.185302} {\bibfield  {journal}
  {\bibinfo  {journal} {Phys. Rev. Lett.}\ }\textbf {\bibinfo {volume} {111}},\
  \bibinfo {pages} {185302} (\bibinfo {year} {2013})}\BibitemShut {NoStop}%
\bibitem [{\citenamefont {Jotzu}\ \emph {et~al.}(2014)\citenamefont {Jotzu},
  \citenamefont {Messer}, \citenamefont {Desbuquois}, \citenamefont {Lebrat},
  \citenamefont {Uehlinger}, \citenamefont {Greif},\ and\ \citenamefont
  {Esslinger}}]{jotzu_experimental_2014}%
  \BibitemOpen
  \bibfield  {author} {\bibinfo {author} {\bibfnamefont {G.}~\bibnamefont
  {Jotzu}}, \bibinfo {author} {\bibfnamefont {M.}~\bibnamefont {Messer}},
  \bibinfo {author} {\bibfnamefont {R.}~\bibnamefont {Desbuquois}}, \bibinfo
  {author} {\bibfnamefont {M.}~\bibnamefont {Lebrat}}, \bibinfo {author}
  {\bibfnamefont {T.}~\bibnamefont {Uehlinger}}, \bibinfo {author}
  {\bibfnamefont {D.}~\bibnamefont {Greif}}, \ and\ \bibinfo {author}
  {\bibfnamefont {T.}~\bibnamefont {Esslinger}},\ }\bibfield  {title} {\enquote
  {\bibinfo {title} {Experimental realization of the topological {Haldane}
  model with ultracold fermions},}\ }\href {\doibase 10.1038/nature13915}
  {\bibfield  {journal} {\bibinfo  {journal} {Nature}\ }\textbf {\bibinfo
  {volume} {515}},\ \bibinfo {pages} {237--240} (\bibinfo {year}
  {2014})}\BibitemShut {NoStop}%
\bibitem [{\citenamefont {Wu}\ \emph {et~al.}(2016)\citenamefont {Wu},
  \citenamefont {Zhang}, \citenamefont {Sun}, \citenamefont {Xu}, \citenamefont
  {Wang}, \citenamefont {Ji}, \citenamefont {Deng}, \citenamefont {Chen},
  \citenamefont {Liu},\ and\ \citenamefont {Pan}}]{wu_realization_2016}%
  \BibitemOpen
  \bibfield  {author} {\bibinfo {author} {\bibfnamefont {Z.}~\bibnamefont
  {Wu}}, \bibinfo {author} {\bibfnamefont {L.}~\bibnamefont {Zhang}}, \bibinfo
  {author} {\bibfnamefont {W.}~\bibnamefont {Sun}}, \bibinfo {author}
  {\bibfnamefont {X.-T.}\ \bibnamefont {Xu}}, \bibinfo {author} {\bibfnamefont
  {B.-Z.}\ \bibnamefont {Wang}}, \bibinfo {author} {\bibfnamefont {S.-C.}\
  \bibnamefont {Ji}}, \bibinfo {author} {\bibfnamefont {Y.}~\bibnamefont
  {Deng}}, \bibinfo {author} {\bibfnamefont {S.}~\bibnamefont {Chen}}, \bibinfo
  {author} {\bibfnamefont {X.-J.}\ \bibnamefont {Liu}}, \ and\ \bibinfo
  {author} {\bibfnamefont {J.-W.}\ \bibnamefont {Pan}},\ }\bibfield  {title}
  {\enquote {\bibinfo {title} {Realization of two-dimensional spin-orbit
  coupling for {Bose}-{Einstein} condensates},}\ }\href {\doibase
  10.1126/science.aaf6689} {\bibfield  {journal} {\bibinfo  {journal}
  {Science}\ }\textbf {\bibinfo {volume} {354}},\ \bibinfo {pages} {83--88}
  (\bibinfo {year} {2016})}\BibitemShut {NoStop}%
\bibitem [{\citenamefont {Thouless}\ \emph {et~al.}(1982)\citenamefont
  {Thouless}, \citenamefont {Kohmoto}, \citenamefont {Nightingale},\ and\
  \citenamefont {den Nijs}}]{thouless_quantized_1982}%
  \BibitemOpen
  \bibfield  {author} {\bibinfo {author} {\bibfnamefont {D.~J.}\ \bibnamefont
  {Thouless}}, \bibinfo {author} {\bibfnamefont {M.}~\bibnamefont {Kohmoto}},
  \bibinfo {author} {\bibfnamefont {M.~P.}\ \bibnamefont {Nightingale}}, \ and\
  \bibinfo {author} {\bibfnamefont {M.}~\bibnamefont {den Nijs}},\ }\bibfield
  {title} {\enquote {\bibinfo {title} {Quantized {Hall} {Conductance} in a
  {Two}-{Dimensional} {Periodic} {Potential}},}\ }\href {\doibase
  10.1103/PhysRevLett.49.405} {\bibfield  {journal} {\bibinfo  {journal} {Phys.
  Rev. Lett.}\ }\textbf {\bibinfo {volume} {49}},\ \bibinfo {pages} {405--408}
  (\bibinfo {year} {1982})}\BibitemShut {NoStop}%
\bibitem [{\citenamefont {Xiao}\ \emph {et~al.}(2010)\citenamefont {Xiao},
  \citenamefont {Chang},\ and\ \citenamefont {Niu}}]{xiao_berry_2010}%
  \BibitemOpen
  \bibfield  {author} {\bibinfo {author} {\bibfnamefont {D.}~\bibnamefont
  {Xiao}}, \bibinfo {author} {\bibfnamefont {M.-C.}\ \bibnamefont {Chang}}, \
  and\ \bibinfo {author} {\bibfnamefont {Q.}~\bibnamefont {Niu}},\ }\bibfield
  {title} {\enquote {\bibinfo {title} {Berry phase effects on electronic
  properties},}\ }\href {\doibase 10.1103/RevModPhys.82.1959} {\bibfield
  {journal} {\bibinfo  {journal} {Rev. Mod. Phys.}\ }\textbf {\bibinfo {volume}
  {82}},\ \bibinfo {pages} {1959--2007} (\bibinfo {year} {2010})}\BibitemShut
  {NoStop}%
\bibitem [{\citenamefont {Aidelsburger}\ \emph {et~al.}(2015)\citenamefont
  {Aidelsburger}, \citenamefont {Lohse}, \citenamefont {Schweizer},
  \citenamefont {Atala}, \citenamefont {Barreiro}, \citenamefont {Nascimb\`ene},
  \citenamefont {Cooper}, \citenamefont {Bloch},\ and\ \citenamefont
  {Goldman}}]{aidelsburger_measuring_2015}%
  \BibitemOpen
  \bibfield  {author} {\bibinfo {author} {\bibfnamefont {M.}~\bibnamefont
  {Aidelsburger}}, \bibinfo {author} {\bibfnamefont {M.}~\bibnamefont {Lohse}},
  \bibinfo {author} {\bibfnamefont {C.}~\bibnamefont {Schweizer}}, \bibinfo
  {author} {\bibfnamefont {M.}~\bibnamefont {Atala}}, \bibinfo {author}
  {\bibfnamefont {J.~T.}\ \bibnamefont {Barreiro}}, \bibinfo {author}
  {\bibfnamefont {S.}~\bibnamefont {Nascimb\`ene}}, \bibinfo {author}
  {\bibfnamefont {N.~R.}\ \bibnamefont {Cooper}}, \bibinfo {author}
  {\bibfnamefont {I.}~\bibnamefont {Bloch}}, \ and\ \bibinfo {author}
  {\bibfnamefont {N.}~\bibnamefont {Goldman}},\ }\bibfield  {title} {\enquote
  {\bibinfo {title} {Measuring the {Chern} number of {Hofstadter} bands with
  ultracold bosonic atoms},}\ }\href {\doibase 10.1038/nphys3171} {\bibfield
  {journal} {\bibinfo  {journal} {Nature Phys.}\ }\textbf {\bibinfo {volume}
  {11}},\ \bibinfo {pages} {162--166} (\bibinfo {year} {2015})}\BibitemShut
  {NoStop}%
\bibitem [{\citenamefont {Mittal}\ \emph {et~al.}(2016)\citenamefont {Mittal},
  \citenamefont {Ganeshan}, \citenamefont {Fan}, \citenamefont {Vaezi},\ and\
  \citenamefont {Hafezi}}]{mittal_measurement_2016}%
  \BibitemOpen
  \bibfield  {author} {\bibinfo {author} {\bibfnamefont {S.}~\bibnamefont
  {Mittal}}, \bibinfo {author} {\bibfnamefont {S.}~\bibnamefont {Ganeshan}},
  \bibinfo {author} {\bibfnamefont {J.}~\bibnamefont {Fan}}, \bibinfo {author}
  {\bibfnamefont {A.}~\bibnamefont {Vaezi}}, \ and\ \bibinfo {author}
  {\bibfnamefont {M.}~\bibnamefont {Hafezi}},\ }\bibfield  {title} {\enquote
  {\bibinfo {title} {Measurement of topological invariants in a {2D} photonic
  system},}\ }\href {\doibase 10.1038/nphoton.2016.10} {\bibfield  {journal}
  {\bibinfo  {journal} {Nature Photon.}\ }\textbf {\bibinfo {volume} {10}},\
  \bibinfo {pages} {180--183} (\bibinfo {year} {2016})}\BibitemShut {NoStop}%
\bibitem [{\citenamefont {Fl\"aschner}\ \emph {et~al.}(2016)\citenamefont
  {Fl\"aschner}, \citenamefont {Rem}, \citenamefont {Tarnowski}, \citenamefont
  {Vogel}, \citenamefont {L\"uhmann}, \citenamefont {Sengstock},\ and\
  \citenamefont {Weitenberg}}]{flaschner_experimental_2016}%
  \BibitemOpen
  \bibfield  {author} {\bibinfo {author} {\bibfnamefont {N.}~\bibnamefont
  {Fl\"aschner}}, \bibinfo {author} {\bibfnamefont {B.~S.}\ \bibnamefont {Rem}},
  \bibinfo {author} {\bibfnamefont {M.}~\bibnamefont {Tarnowski}}, \bibinfo
  {author} {\bibfnamefont {D.}~\bibnamefont {Vogel}}, \bibinfo {author}
  {\bibfnamefont {D.-S.}\ \bibnamefont {L\"uhmann}}, \bibinfo {author}
  {\bibfnamefont {K.}~\bibnamefont {Sengstock}}, \ and\ \bibinfo {author}
  {\bibfnamefont {Christof}\ \bibnamefont {Weitenberg}},\ }\bibfield  {title}
  {\enquote {\bibinfo {title} {Experimental reconstruction of the {Berry}
  curvature in a {Floquet} {Bloch} band},}\ }\href
  {https://science.sciencemag.org/content/352/6289/1091.abstract} {\bibfield
  {journal} {\bibinfo  {journal} {Science}\ }\textbf {\bibinfo {volume}
  {352}},\ \bibinfo {pages} {1091--1094} (\bibinfo {year} {2016})}\BibitemShut
  {NoStop}%
\bibitem [{\citenamefont {Tarnowski}\ \emph {et~al.}(2019)\citenamefont
  {Tarnowski}, \citenamefont {\"Unal}, \citenamefont {Fl\"aschner}, \citenamefont
  {Rem}, \citenamefont {Eckardt}, \citenamefont {Sengstock},\ and\
  \citenamefont {Weitenberg}}]{tarnowski_measuring_2019}%
  \BibitemOpen
  \bibfield  {author} {\bibinfo {author} {\bibfnamefont {M.}~\bibnamefont
  {Tarnowski}}, \bibinfo {author} {\bibfnamefont {F.~N.}\ \bibnamefont
  {\"Unal}}, \bibinfo {author} {\bibfnamefont {N.}~\bibnamefont {Fl\"aschner}},
  \bibinfo {author} {\bibfnamefont {B.~S.}\ \bibnamefont {Rem}}, \bibinfo
  {author} {\bibfnamefont {A.}~\bibnamefont {Eckardt}}, \bibinfo {author}
  {\bibfnamefont {K.}~\bibnamefont {Sengstock}}, \ and\ \bibinfo {author}
  {\bibfnamefont {C.}~\bibnamefont {Weitenberg}},\ }\bibfield  {title}
  {\enquote {\bibinfo {title} {Measuring topology from dynamics by obtaining
  the {Chern} number from a linking number},}\ }\href {\doibase
  10.1038/s41467-019-09668-y} {\bibfield  {journal} {\bibinfo  {journal}
  {Nature Commun.}\ }\textbf {\bibinfo {volume} {10}},\ \bibinfo {pages}
  {1--13} (\bibinfo {year} {2019})}\BibitemShut {NoStop}%
\bibitem [{\citenamefont {Asteria}\ \emph {et~al.}(2019)\citenamefont
  {Asteria}, \citenamefont {Thanh~Tran}, \citenamefont {Ozawa}, \citenamefont
  {Tarnowski}, \citenamefont {Rem}, \citenamefont {Fl\"aschner}, \citenamefont
  {Sengstock}, \citenamefont {Goldman},\ and\ \citenamefont
  {Weitenberg}}]{asteria_measuring_2019}%
  \BibitemOpen
  \bibfield  {author} {\bibinfo {author} {\bibfnamefont {L.}~\bibnamefont
  {Asteria}}, \bibinfo {author} {\bibfnamefont {D.}~\bibnamefont {Thanh~Tran}},
  \bibinfo {author} {\bibfnamefont {T.}~\bibnamefont {Ozawa}}, \bibinfo
  {author} {\bibfnamefont {M.}~\bibnamefont {Tarnowski}}, \bibinfo {author}
  {\bibfnamefont {B.~S.}\ \bibnamefont {Rem}}, \bibinfo {author} {\bibfnamefont
  {N.}~\bibnamefont {Fl\"aschner}}, \bibinfo {author} {\bibfnamefont
  {K.}~\bibnamefont {Sengstock}}, \bibinfo {author} {\bibfnamefont
  {N.}~\bibnamefont {Goldman}}, \ and\ \bibinfo {author} {\bibfnamefont
  {C.}~\bibnamefont {Weitenberg}},\ }\bibfield  {title} {\enquote {\bibinfo
  {title} {Measuring quantized circular dichroism in ultracold topological
  matter},}\ }\href {https://www.nature.com/articles/s41567-019-0417-8}
  {\bibfield  {journal} {\bibinfo  {journal} {Nature Phys.}\ }\textbf {\bibinfo
  {volume} {15}},\ \bibinfo {pages} {449--454} (\bibinfo {year}
  {2019})}\BibitemShut {NoStop}%
\bibitem [{\citenamefont {Hatsugai}(1993)}]{hatsugai_chern_1993}%
  \BibitemOpen
  \bibfield  {author} {\bibinfo {author} {\bibfnamefont {Y.}~\bibnamefont
  {Hatsugai}},\ }\bibfield  {title} {\enquote {\bibinfo {title} {Chern number
  and edge states in the integer quantum {Hall} effect},}\ }\href {\doibase
  10.1103/PhysRevLett.71.3697} {\bibfield  {journal} {\bibinfo  {journal}
  {Phys. Rev. Lett.}\ }\textbf {\bibinfo {volume} {71}},\ \bibinfo {pages}
  {3697--3700} (\bibinfo {year} {1993})}\BibitemShut {NoStop}%
\bibitem [{\citenamefont {Qi}\ \emph {et~al.}(2006)\citenamefont {Qi},
  \citenamefont {Wu},\ and\ \citenamefont {Zhang}}]{qi_general_2006}%
  \BibitemOpen
  \bibfield  {author} {\bibinfo {author} {\bibfnamefont {X.-L.}\ \bibnamefont
  {Qi}}, \bibinfo {author} {\bibfnamefont {Y.-S.}\ \bibnamefont {Wu}}, \ and\
  \bibinfo {author} {\bibfnamefont {S.-C.}\ \bibnamefont {Zhang}},\ }\bibfield
  {title} {\enquote {\bibinfo {title} {General theorem relating the bulk
  topological number to edge states in two-dimensional insulators},}\ }\href
  {\doibase 10.1103/PhysRevB.74.045125} {\bibfield  {journal} {\bibinfo
  {journal} {Phys. Rev. B}\ }\textbf {\bibinfo {volume} {74}},\ \bibinfo
  {pages} {045125} (\bibinfo {year} {2006})}\BibitemShut {NoStop}%
\bibitem [{\citenamefont {Kitagawa}\ \emph {et~al.}(2010)\citenamefont
  {Kitagawa}, \citenamefont {Berg}, \citenamefont {Rudner},\ and\ \citenamefont
  {Demler}}]{kitagawa_topological_2010}%
  \BibitemOpen
  \bibfield  {author} {\bibinfo {author} {\bibfnamefont {T.}~\bibnamefont
  {Kitagawa}}, \bibinfo {author} {\bibfnamefont {E.}~\bibnamefont {Berg}},
  \bibinfo {author} {\bibfnamefont {M.}~\bibnamefont {Rudner}}, \ and\ \bibinfo
  {author} {\bibfnamefont {E.}~\bibnamefont {Demler}},\ }\bibfield  {title}
  {\enquote {\bibinfo {title} {Topological characterization of periodically
  driven quantum systems},}\ }\href {\doibase 10.1103/PhysRevB.82.235114}
  {\bibfield  {journal} {\bibinfo  {journal} {Phys. Rev. B}\ }\textbf {\bibinfo
  {volume} {82}},\ \bibinfo {pages} {235114} (\bibinfo {year}
  {2010})}\BibitemShut {NoStop}%
\bibitem [{\citenamefont {Rudner}\ \emph {et~al.}(2013)\citenamefont {Rudner},
  \citenamefont {Lindner}, \citenamefont {Berg},\ and\ \citenamefont
  {Levin}}]{rudner_anomalous_2013}%
  \BibitemOpen
  \bibfield  {author} {\bibinfo {author} {\bibfnamefont {M.~S.}\ \bibnamefont
  {Rudner}}, \bibinfo {author} {\bibfnamefont {N.~H.}\ \bibnamefont {Lindner}},
  \bibinfo {author} {\bibfnamefont {E.}~\bibnamefont {Berg}}, \ and\ \bibinfo
  {author} {\bibfnamefont {M.}~\bibnamefont {Levin}},\ }\bibfield  {title}
  {\enquote {\bibinfo {title} {Anomalous {Edge} {States} and the {Bulk}-{Edge}
  {Correspondence} for {Periodically} {Driven} {Two}-{Dimensional}
  {Systems}},}\ }\href {\doibase 10.1103/PhysRevX.3.031005} {\bibfield
  {journal} {\bibinfo  {journal} {Phys. Rev. X}\ }\textbf {\bibinfo {volume}
  {3}},\ \bibinfo {pages} {031005} (\bibinfo {year} {2013})}\BibitemShut
  {NoStop}%
\bibitem [{\citenamefont {Nathan}\ and\ \citenamefont
  {Rudner}(2015)}]{nathan_topological_2015}%
  \BibitemOpen
  \bibfield  {author} {\bibinfo {author} {\bibfnamefont {F.}~\bibnamefont
  {Nathan}}\ and\ \bibinfo {author} {\bibfnamefont {M.~S.}\ \bibnamefont
  {Rudner}},\ }\bibfield  {title} {\enquote {\bibinfo {title} {Topological
  singularities and the general classification of {Floquet}-{Bloch} systems},}\
  }\href {\doibase 10.1088/1367-2630/17/12/125014} {\bibfield  {journal}
  {\bibinfo  {journal} {New J. Phys.}\ }\textbf {\bibinfo {volume} {17}},\
  \bibinfo {pages} {125014} (\bibinfo {year} {2015})}\BibitemShut {NoStop}%
\bibitem [{\citenamefont {Rudner}\ and\ \citenamefont
  {Lindner}(2020)}]{rudner_floquet_2019}%
  \BibitemOpen
  \bibfield  {author} {\bibinfo {author} {\bibfnamefont {M.~S.}\ \bibnamefont
  {Rudner}}\ and\ \bibinfo {author} {\bibfnamefont {N.~H.}\ \bibnamefont
  {Lindner}},\ }\bibfield  {title} {\enquote {\bibinfo {title} {Band structure
  engineering and non-equilibrium dynamics in floquet topological
  insulators},}\ }\href {https://www.nature.com/articles/s42254-020-0170-z}
  {\bibfield  {journal} {\bibinfo  {journal} {Nat. Rev. Phys.}\ }\textbf
  {\bibinfo {volume} {2}},\ \bibinfo {pages} {229--244} (\bibinfo {year}
  {2020})}\BibitemShut {NoStop}%
\bibitem [{\citenamefont {Nathan}\ \emph {et~al.}(2019)\citenamefont {Nathan},
  \citenamefont {Abanin}, \citenamefont {Berg}, \citenamefont {Lindner},\ and\
  \citenamefont {Rudner}}]{nathan_anomalous_2019}%
  \BibitemOpen
  \bibfield  {author} {\bibinfo {author} {\bibfnamefont {F.}~\bibnamefont
  {Nathan}}, \bibinfo {author} {\bibfnamefont {D.}~\bibnamefont {Abanin}},
  \bibinfo {author} {\bibfnamefont {E.}~\bibnamefont {Berg}}, \bibinfo {author}
  {\bibfnamefont {N.~H.}\ \bibnamefont {Lindner}}, \ and\ \bibinfo {author}
  {\bibfnamefont {M.~S.}\ \bibnamefont {Rudner}},\ }\bibfield  {title}
  {\enquote {\bibinfo {title} {Anomalous {Floquet} insulators},}\ }\href
  {\doibase 10.1103/PhysRevB.99.195133} {\bibfield  {journal} {\bibinfo
  {journal} {Phys. Rev. B}\ }\textbf {\bibinfo {volume} {99}},\ \bibinfo
  {pages} {195133} (\bibinfo {year} {2019})}\BibitemShut {NoStop}%
\bibitem [{\citenamefont {Kitagawa}\ \emph {et~al.}(2012)\citenamefont
  {Kitagawa}, \citenamefont {Broome}, \citenamefont {Fedrizzi}, \citenamefont
  {Rudner}, \citenamefont {Berg}, \citenamefont {Kassal}, \citenamefont
  {Aspuru-Guzik}, \citenamefont {Demler},\ and\ \citenamefont
  {White}}]{kitagawa_observation_2012}%
  \BibitemOpen
  \bibfield  {author} {\bibinfo {author} {\bibfnamefont {T.}~\bibnamefont
  {Kitagawa}}, \bibinfo {author} {\bibfnamefont {M.~A.}\ \bibnamefont
  {Broome}}, \bibinfo {author} {\bibfnamefont {A.}~\bibnamefont {Fedrizzi}},
  \bibinfo {author} {\bibfnamefont {M.~S.}\ \bibnamefont {Rudner}}, \bibinfo
  {author} {\bibfnamefont {E.}~\bibnamefont {Berg}}, \bibinfo {author}
  {\bibfnamefont {I.}~\bibnamefont {Kassal}}, \bibinfo {author} {\bibfnamefont
  {A.}~\bibnamefont {Aspuru-Guzik}}, \bibinfo {author} {\bibfnamefont
  {E.}~\bibnamefont {Demler}}, \ and\ \bibinfo {author} {\bibfnamefont {A.~G.}\
  \bibnamefont {White}},\ }\bibfield  {title} {\enquote {\bibinfo {title}
  {Observation of topologically protected bound states in photonic quantum
  walks},}\ }\href {\doibase 10.1038/ncomms1872} {\bibfield  {journal}
  {\bibinfo  {journal} {Nature Commun.}\ }\textbf {\bibinfo {volume} {3}},\
  \bibinfo {pages} {1--7} (\bibinfo {year} {2012})}\BibitemShut {NoStop}%
\bibitem [{\citenamefont {Hu}\ \emph {et~al.}(2015)\citenamefont {Hu},
  \citenamefont {Pillay}, \citenamefont {Wu}, \citenamefont {Pasek},
  \citenamefont {Shum},\ and\ \citenamefont {Chong}}]{hu_measurement_2015}%
  \BibitemOpen
  \bibfield  {author} {\bibinfo {author} {\bibfnamefont {W.}~\bibnamefont
  {Hu}}, \bibinfo {author} {\bibfnamefont {J.~C.}\ \bibnamefont {Pillay}},
  \bibinfo {author} {\bibfnamefont {K.}~\bibnamefont {Wu}}, \bibinfo {author}
  {\bibfnamefont {M.}~\bibnamefont {Pasek}}, \bibinfo {author} {\bibfnamefont
  {P.~P.}\ \bibnamefont {Shum}}, \ and\ \bibinfo {author} {\bibfnamefont
  {Y.~D.}\ \bibnamefont {Chong}},\ }\bibfield  {title} {\enquote {\bibinfo
  {title} {Measurement of a {Topological} {Edge} {Invariant} in a {Microwave}
  {Network}},}\ }\href {\doibase 10.1103/PhysRevX.5.011012} {\bibfield
  {journal} {\bibinfo  {journal} {Phys. Rev. X}\ }\textbf {\bibinfo {volume}
  {5}},\ \bibinfo {pages} {011012} (\bibinfo {year} {2015})}\BibitemShut
  {NoStop}%
\bibitem [{\citenamefont {Maczewsky}\ \emph {et~al.}(2017)\citenamefont
  {Maczewsky}, \citenamefont {Zeuner}, \citenamefont {Nolte},\ and\
  \citenamefont {Szameit}}]{maczewsky_observation_2017}%
  \BibitemOpen
  \bibfield  {author} {\bibinfo {author} {\bibfnamefont {L.~J.}\ \bibnamefont
  {Maczewsky}}, \bibinfo {author} {\bibfnamefont {J.~M.}\ \bibnamefont
  {Zeuner}}, \bibinfo {author} {\bibfnamefont {S.}~\bibnamefont {Nolte}}, \
  and\ \bibinfo {author} {\bibfnamefont {A.}~\bibnamefont {Szameit}},\
  }\bibfield  {title} {\enquote {\bibinfo {title} {Observation of photonic
  anomalous {Floquet} topological insulators},}\ }\href
  {https://www.nature.com/articles/ncomms13756} {\bibfield  {journal} {\bibinfo
   {journal} {Nature Commun.}\ }\textbf {\bibinfo {volume} {8}},\ \bibinfo
  {pages} {13756} (\bibinfo {year} {2017})}\BibitemShut {NoStop}%
\bibitem [{\citenamefont {Mukherjee}\ \emph {et~al.}(2017)\citenamefont
  {Mukherjee}, \citenamefont {Spracklen}, \citenamefont {Valiente},
  \citenamefont {Andersson}, \citenamefont {\"Ohberg}, \citenamefont {Goldman},\
  and\ \citenamefont {Thomson}}]{mukherjee_experimental_2017}%
  \BibitemOpen
  \bibfield  {author} {\bibinfo {author} {\bibfnamefont {S.}~\bibnamefont
  {Mukherjee}}, \bibinfo {author} {\bibfnamefont {A.}~\bibnamefont
  {Spracklen}}, \bibinfo {author} {\bibfnamefont {M.}~\bibnamefont {Valiente}},
  \bibinfo {author} {\bibfnamefont {E.}~\bibnamefont {Andersson}}, \bibinfo
  {author} {\bibfnamefont {P.}~\bibnamefont {\"Ohberg}}, \bibinfo {author}
  {\bibfnamefont {N.}~\bibnamefont {Goldman}}, \ and\ \bibinfo {author}
  {\bibfnamefont {R.~R.}\ \bibnamefont {Thomson}},\ }\bibfield  {title}
  {\enquote {\bibinfo {title} {Experimental observation of anomalous
  topological edge modes in a slowly driven photonic lattice},}\ }\href
  {\doibase 10.1038/ncomms13918} {\bibfield  {journal} {\bibinfo  {journal}
  {Nature Commun.}\ }\textbf {\bibinfo {volume} {8}},\ \bibinfo {pages} {1--7}
  (\bibinfo {year} {2017})}\BibitemShut {NoStop}%
\bibitem [{\citenamefont {Peng}\ \emph {et~al.}(2016)\citenamefont {Peng},
  \citenamefont {Qin}, \citenamefont {Zhao}, \citenamefont {Shen},
  \citenamefont {Xu}, \citenamefont {Bao}, \citenamefont {Jia},\ and\
  \citenamefont {Zhu}}]{peng_experimental_2016}%
  \BibitemOpen
  \bibfield  {author} {\bibinfo {author} {\bibfnamefont {Y.-G.}\ \bibnamefont
  {Peng}}, \bibinfo {author} {\bibfnamefont {C.-Z.}\ \bibnamefont {Qin}},
  \bibinfo {author} {\bibfnamefont {D.-G.}\ \bibnamefont {Zhao}}, \bibinfo
  {author} {\bibfnamefont {Y.-X.}\ \bibnamefont {Shen}}, \bibinfo {author}
  {\bibfnamefont {X.-Y.}\ \bibnamefont {Xu}}, \bibinfo {author} {\bibfnamefont
  {M.}~\bibnamefont {Bao}}, \bibinfo {author} {\bibfnamefont {H.}~\bibnamefont
  {Jia}}, \ and\ \bibinfo {author} {\bibfnamefont {X.-F.}\ \bibnamefont
  {Zhu}},\ }\bibfield  {title} {\enquote {\bibinfo {title} {Experimental
  demonstration of anomalous {Floquet} topological insulator for sound},}\
  }\href {\doibase 10.1038/ncomms13368} {\bibfield  {journal} {\bibinfo
  {journal} {Nat. Commun.}\ }\textbf {\bibinfo {volume} {7}},\ \bibinfo {pages}
  {1--8} (\bibinfo {year} {2016})}\BibitemShut {NoStop}%
\bibitem [{\citenamefont {Zenesini}\ \emph {et~al.}(2010)\citenamefont
  {Zenesini}, \citenamefont {Ciampini}, \citenamefont {Morsch},\ and\
  \citenamefont {Arimondo}}]{zenesini_observation_2010}%
  \BibitemOpen
  \bibfield  {author} {\bibinfo {author} {\bibfnamefont {A.}~\bibnamefont
  {Zenesini}}, \bibinfo {author} {\bibfnamefont {D.}~\bibnamefont {Ciampini}},
  \bibinfo {author} {\bibfnamefont {O.}~\bibnamefont {Morsch}}, \ and\ \bibinfo
  {author} {\bibfnamefont {E.}~\bibnamefont {Arimondo}},\ }\bibfield  {title}
  {\enquote {\bibinfo {title} {Observation of {St\"uckelberg} oscillations in
  accelerated optical lattices},}\ }\href {\doibase 10.1103/PhysRevA.82.065601}
  {\bibfield  {journal} {\bibinfo  {journal} {Phys. Rev. A}\ }\textbf {\bibinfo
  {volume} {82}},\ \bibinfo {pages} {065601} (\bibinfo {year}
  {2010})}\BibitemShut {NoStop}%
\bibitem [{\citenamefont {Kling}\ \emph {et~al.}(2010)\citenamefont {Kling},
  \citenamefont {Salger}, \citenamefont {Grossert},\ and\ \citenamefont
  {Weitz}}]{kling_atomic_2010}%
  \BibitemOpen
  \bibfield  {author} {\bibinfo {author} {\bibfnamefont {S.}~\bibnamefont
  {Kling}}, \bibinfo {author} {\bibfnamefont {T.}~\bibnamefont {Salger}},
  \bibinfo {author} {\bibfnamefont {C.}~\bibnamefont {Grossert}}, \ and\
  \bibinfo {author} {\bibfnamefont {M.}~\bibnamefont {Weitz}},\ }\bibfield
  {title} {\enquote {\bibinfo {title} {Atomic {Bloch}-{Zener} {Oscillations}
  and {St\"uckelberg} {Interferometry} in {Optical} {Lattices}},}\ }\href
  {\doibase 10.1103/PhysRevLett.105.215301} {\bibfield  {journal} {\bibinfo
  {journal} {Phys. Rev. Lett.}\ }\textbf {\bibinfo {volume} {105}},\ \bibinfo
  {pages} {215301} (\bibinfo {year} {2010})}\BibitemShut {NoStop}%
\bibitem [{\citenamefont {Quelle}\ \emph {et~al.}(2017)\citenamefont {Quelle},
  \citenamefont {Weitenberg}, \citenamefont {Sengstock},\ and\ \citenamefont
  {Smith}}]{quelle_driving_2017}%
  \BibitemOpen
  \bibfield  {author} {\bibinfo {author} {\bibfnamefont {A.}~\bibnamefont
  {Quelle}}, \bibinfo {author} {\bibfnamefont {C.}~\bibnamefont {Weitenberg}},
  \bibinfo {author} {\bibfnamefont {K.}~\bibnamefont {Sengstock}}, \ and\
  \bibinfo {author} {\bibfnamefont {C.~Morais}\ \bibnamefont {Smith}},\
  }\bibfield  {title} {\enquote {\bibinfo {title} {Driving protocol for a
  {Floquet} topological phase without static counterpart},}\ }\href {\doibase
  10.1088/1367-2630/aa8646} {\bibfield  {journal} {\bibinfo  {journal} {New J.
  Phys.}\ }\textbf {\bibinfo {volume} {19}},\ \bibinfo {pages} {113010}
  (\bibinfo {year} {2017})}\BibitemShut {NoStop}%
\bibitem [{\citenamefont {\"Unal}\ \emph {et~al.}(2019)\citenamefont {\"Unal},
  \citenamefont {Seradjeh},\ and\ \citenamefont {Eckardt}}]{unal_how_2019}%
  \BibitemOpen
  \bibfield  {author} {\bibinfo {author} {\bibfnamefont {F.~N.}\ \bibnamefont
  {\"Unal}}, \bibinfo {author} {\bibfnamefont {B.}~\bibnamefont {Seradjeh}}, \
  and\ \bibinfo {author} {\bibfnamefont {A.}~\bibnamefont {Eckardt}},\
  }\bibfield  {title} {\enquote {\bibinfo {title} {How to {Directly} {Measure}
  {Floquet} {Topological} {Invariants} in {Optical} {Lattices}},}\ }\href
  {\doibase 10.1103/PhysRevLett.122.253601} {\bibfield  {journal} {\bibinfo
  {journal} {Phys. Rev. Lett.}\ }\textbf {\bibinfo {volume} {122}},\ \bibinfo
  {pages} {253601} (\bibinfo {year} {2019})}\BibitemShut {NoStop}%
\bibitem [{\citenamefont {Greiner}\ \emph {et~al.}(2001)\citenamefont
  {Greiner}, \citenamefont {Bloch}, \citenamefont {Mandel}, \citenamefont
  {H\"ansch},\ and\ \citenamefont {Esslinger}}]{greiner_exploring_2001}%
  \BibitemOpen
  \bibfield  {author} {\bibinfo {author} {\bibfnamefont {M.}~\bibnamefont
  {Greiner}}, \bibinfo {author} {\bibfnamefont {I.}~\bibnamefont {Bloch}},
  \bibinfo {author} {\bibfnamefont {O.}~\bibnamefont {Mandel}}, \bibinfo
  {author} {\bibfnamefont {T.~W.}\ \bibnamefont {H\"ansch}}, \ and\ \bibinfo
  {author} {\bibfnamefont {T.}~\bibnamefont {Esslinger}},\ }\bibfield  {title}
  {\enquote {\bibinfo {title} {Exploring {Phase} {Coherence} in a {2D}
  {Lattice} of {Bose}-{Einstein} {Condensates}},}\ }\href {\doibase
  10.1103/PhysRevLett.87.160405} {\bibfield  {journal} {\bibinfo  {journal}
  {Phys. Rev. Lett.}\ }\textbf {\bibinfo {volume} {87}},\ \bibinfo {pages}
  {160405} (\bibinfo {year} {2001})}\BibitemShut {NoStop}%
\bibitem [{\citenamefont {Bouhon}\ \emph {et~al.}(2019)\citenamefont {Bouhon},
  \citenamefont {Black-Schaffer},\ and\ \citenamefont
  {Slager}}]{bouhon_wilson_2019}%
  \BibitemOpen
  \bibfield  {author} {\bibinfo {author} {\bibfnamefont {A.}~\bibnamefont
  {Bouhon}}, \bibinfo {author} {\bibfnamefont {A.~M.}\ \bibnamefont
  {Black-Schaffer}}, \ and\ \bibinfo {author} {\bibfnamefont {R.-J.}\
  \bibnamefont {Slager}},\ }\bibfield  {title} {\enquote {\bibinfo {title}
  {Wilson loop approach to fragile topology of split elementary band
  representations and topological crystalline insulators with time-reversal
  symmetry},}\ }\href {\doibase 10.1103/PhysRevB.100.195135} {\bibfield
  {journal} {\bibinfo  {journal} {Phys. Rev. B}\ }\textbf {\bibinfo {volume}
  {100}},\ \bibinfo {pages} {195135} (\bibinfo {year} {2019})}\BibitemShut
  {NoStop}%
\bibitem [{\citenamefont {Simon}(1983)}]{simon_holonomy_1983}%
  \BibitemOpen
  \bibfield  {author} {\bibinfo {author} {\bibfnamefont {B.}~\bibnamefont
  {Simon}},\ }\bibfield  {title} {\enquote {\bibinfo {title} {Holonomy, the
  {Quantum} {Adiabatic} {Theorem}, and {Berry}'s {Phase}},}\ }\href {\doibase
  10.1103/PhysRevLett.51.2167} {\bibfield  {journal} {\bibinfo  {journal}
  {Phys. Rev. Lett.}\ }\textbf {\bibinfo {volume} {51}},\ \bibinfo {pages}
  {2167--2170} (\bibinfo {year} {1983})}\BibitemShut {NoStop}%
\bibitem [{\citenamefont {Bellissard}(1995)}]{bellissard_change_1995}%
  \BibitemOpen
  \bibfield  {author} {\bibinfo {author} {\bibfnamefont {J.}~\bibnamefont
  {Bellissard}},\ }\bibfield  {title} {\enquote {\bibinfo {title} {Change of
  the {Chern} number at band crossings},}\ }\href
  {http://arxiv.org/abs/cond-mat/9504030} {\bibfield  {journal} {\bibinfo
  {journal} {arXiv}\ }\textbf {\bibinfo {volume} {cond-mat/9504030}} (\bibinfo
  {year} {1995})}\BibitemShut {NoStop}%
\bibitem [{\citenamefont {Leboeuf}\ \emph {et~al.}(1992)\citenamefont
  {Leboeuf}, \citenamefont {Kurchan}, \citenamefont {Feingold},\ and\
  \citenamefont {Arovas}}]{leboeuf_topological_1992}%
  \BibitemOpen
  \bibfield  {author} {\bibinfo {author} {\bibfnamefont {P.}~\bibnamefont
  {Leboeuf}}, \bibinfo {author} {\bibfnamefont {J.}~\bibnamefont {Kurchan}},
  \bibinfo {author} {\bibfnamefont {M.}~\bibnamefont {Feingold}}, \ and\
  \bibinfo {author} {\bibfnamefont {D.~P.}\ \bibnamefont {Arovas}},\ }\bibfield
   {title} {\enquote {\bibinfo {title} {Topological aspects of quantum
  chaos},}\ }\href {\doibase 10.1063/1.165915} {\bibfield  {journal} {\bibinfo
  {journal} {Chaos}\ }\textbf {\bibinfo {volume} {2}},\ \bibinfo {pages}
  {125--130} (\bibinfo {year} {1992})}\BibitemShut {NoStop}%
\bibitem [{\citenamefont {Barelli}\ and\ \citenamefont
  {Fleckinger}(1992)}]{barelli_semiclassical_1992}%
  \BibitemOpen
  \bibfield  {author} {\bibinfo {author} {\bibfnamefont {A.}~\bibnamefont
  {Barelli}}\ and\ \bibinfo {author} {\bibfnamefont {R.}~\bibnamefont
  {Fleckinger}},\ }\bibfield  {title} {\enquote {\bibinfo {title}
  {Semiclassical analysis of {Harper}-like models},}\ }\href {\doibase
  10.1103/PhysRevB.46.11559} {\bibfield  {journal} {\bibinfo  {journal} {Phys.
  Rev. B}\ }\textbf {\bibinfo {volume} {46}},\ \bibinfo {pages} {11559--11569}
  (\bibinfo {year} {1992})}\BibitemShut {NoStop}%
\bibitem [{\citenamefont {Oka}\ and\ \citenamefont
  {Aoki}(2009)}]{oka_photovoltaic_2009}%
  \BibitemOpen
  \bibfield  {author} {\bibinfo {author} {\bibfnamefont {T.}~\bibnamefont
  {Oka}}\ and\ \bibinfo {author} {\bibfnamefont {H.}~\bibnamefont {Aoki}},\
  }\bibfield  {title} {\enquote {\bibinfo {title} {Photovoltaic {Hall} effect
  in graphene},}\ }\href {\doibase 10.1103/PhysRevB.79.081406} {\bibfield
  {journal} {\bibinfo  {journal} {Phys. Rev. B}\ }\textbf {\bibinfo {volume}
  {79}},\ \bibinfo {pages} {081406} (\bibinfo {year} {2009})}\BibitemShut
  {NoStop}%
\bibitem [{\citenamefont {Price}\ and\ \citenamefont
  {Cooper}(2012)}]{price_mapping_2012}%
  \BibitemOpen
  \bibfield  {author} {\bibinfo {author} {\bibfnamefont {H.~M.}\ \bibnamefont
  {Price}}\ and\ \bibinfo {author} {\bibfnamefont {N.~R.}\ \bibnamefont
  {Cooper}},\ }\bibfield  {title} {\enquote {\bibinfo {title} {Mapping the
  {Berry} curvature from semiclassical dynamics in optical lattices},}\ }\href
  {\doibase 10.1103/PhysRevA.85.033620} {\bibfield  {journal} {\bibinfo
  {journal} {Phys. Rev. A}\ }\textbf {\bibinfo {volume} {85}},\ \bibinfo
  {pages} {033620} (\bibinfo {year} {2012})}\BibitemShut {NoStop}%
\bibitem [{\citenamefont {Dauphin}\ and\ \citenamefont
  {Goldman}(2013)}]{dauphin_extracting_2013}%
  \BibitemOpen
  \bibfield  {author} {\bibinfo {author} {\bibfnamefont {A.}~\bibnamefont
  {Dauphin}}\ and\ \bibinfo {author} {\bibfnamefont {N.}~\bibnamefont
  {Goldman}},\ }\bibfield  {title} {\enquote {\bibinfo {title} {Extracting the
  {Chern} {Number} from the {Dynamics} of a {Fermi} {Gas}: {Implementing} a
  {Quantum} {Hall} {Bar} for {Cold} {Atoms}},}\ }\href {\doibase
  10.1103/PhysRevLett.111.135302} {\bibfield  {journal} {\bibinfo  {journal}
  {Phys. Rev. Lett.}\ }\textbf {\bibinfo {volume} {111}},\ \bibinfo {pages}
  {135302} (\bibinfo {year} {2013})}\BibitemShut {NoStop}%
\bibitem [{\citenamefont {Buchhold}\ \emph {et~al.}(2012)\citenamefont
  {Buchhold}, \citenamefont {Cocks},\ and\ \citenamefont
  {Hofstetter}}]{buchhold_effects_2012}%
  \BibitemOpen
  \bibfield  {author} {\bibinfo {author} {\bibfnamefont {M.}~\bibnamefont
  {Buchhold}}, \bibinfo {author} {\bibfnamefont {D.}~\bibnamefont {Cocks}}, \
  and\ \bibinfo {author} {\bibfnamefont {W.}~\bibnamefont {Hofstetter}},\
  }\bibfield  {title} {\enquote {\bibinfo {title} {Effects of smooth boundaries
  on topological edge modes in optical lattices},}\ }\href {\doibase
  10.1103/PhysRevA.85.063614} {\bibfield  {journal} {\bibinfo  {journal} {Phys.
  Rev. A}\ }\textbf {\bibinfo {volume} {85}},\ \bibinfo {pages} {063614}
  (\bibinfo {year} {2012})}\BibitemShut {NoStop}%
\bibitem [{\citenamefont {Goldman}\ \emph {et~al.}(2013)\citenamefont
  {Goldman}, \citenamefont {Dalibard}, \citenamefont {Dauphin}, \citenamefont
  {Gerbier}, \citenamefont {Lewenstein}, \citenamefont {Zoller},\ and\
  \citenamefont {Spielman}}]{goldman_direct_2013}%
  \BibitemOpen
  \bibfield  {author} {\bibinfo {author} {\bibfnamefont {N.}~\bibnamefont
  {Goldman}}, \bibinfo {author} {\bibfnamefont {J.}~\bibnamefont {Dalibard}},
  \bibinfo {author} {\bibfnamefont {A.}~\bibnamefont {Dauphin}}, \bibinfo
  {author} {\bibfnamefont {F.}~\bibnamefont {Gerbier}}, \bibinfo {author}
  {\bibfnamefont {M.}~\bibnamefont {Lewenstein}}, \bibinfo {author}
  {\bibfnamefont {P.}~\bibnamefont {Zoller}}, \ and\ \bibinfo {author}
  {\bibfnamefont {I.~B.}\ \bibnamefont {Spielman}},\ }\bibfield  {title}
  {\enquote {\bibinfo {title} {Direct imaging of topological edge states in
  cold-atom systems},}\ }\href {https://www.pnas.org/content/110/17/6736}
  {\bibfield  {journal} {\bibinfo  {journal} {PNAS}\ }\textbf {\bibinfo
  {volume} {11}},\ \bibinfo {pages} {6736--6741} (\bibinfo {year}
  {2013})}\BibitemShut {NoStop}%
\bibitem [{\citenamefont {Reichl}\ and\ \citenamefont
  {Mueller}(2014)}]{reichl_floquet_2014}%
  \BibitemOpen
  \bibfield  {author} {\bibinfo {author} {\bibfnamefont {M.~D.}\ \bibnamefont
  {Reichl}}\ and\ \bibinfo {author} {\bibfnamefont {E.~J.}\ \bibnamefont
  {Mueller}},\ }\bibfield  {title} {\enquote {\bibinfo {title} {Floquet edge
  states with ultracold atoms},}\ }\href {\doibase 10.1103/PhysRevA.89.063628}
  {\bibfield  {journal} {\bibinfo  {journal} {Phys. Rev. A}\ }\textbf {\bibinfo
  {volume} {89}},\ \bibinfo {pages} {063628} (\bibinfo {year}
  {2014})}\BibitemShut {NoStop}%
\bibitem [{\citenamefont {Titum}\ \emph {et~al.}(2015)\citenamefont {Titum},
  \citenamefont {Lindner}, \citenamefont {Rechtsman},\ and\ \citenamefont
  {Refael}}]{titum_disorder-induced_2015}%
  \BibitemOpen
  \bibfield  {author} {\bibinfo {author} {\bibfnamefont {P.}~\bibnamefont
  {Titum}}, \bibinfo {author} {\bibfnamefont {N.~H.}\ \bibnamefont {Lindner}},
  \bibinfo {author} {\bibfnamefont {M.~C.}\ \bibnamefont {Rechtsman}}, \ and\
  \bibinfo {author} {\bibfnamefont {G.}~\bibnamefont {Refael}},\ }\bibfield
  {title} {\enquote {\bibinfo {title} {Disorder-{Induced} {Floquet}
  {Topological} {Insulators}},}\ }\href {\doibase
  10.1103/PhysRevLett.114.056801} {\bibfield  {journal} {\bibinfo  {journal}
  {Phys. Rev. Lett.}\ }\textbf {\bibinfo {volume} {114}},\ \bibinfo {pages}
  {056801} (\bibinfo {year} {2015})}\BibitemShut {NoStop}%
\bibitem [{\citenamefont {Titum}\ \emph {et~al.}(2016)\citenamefont {Titum},
  \citenamefont {Berg}, \citenamefont {Rudner}, \citenamefont {Refael},\ and\
  \citenamefont {Lindner}}]{titum_anomalous_2016}%
  \BibitemOpen
  \bibfield  {author} {\bibinfo {author} {\bibfnamefont {P.}~\bibnamefont
  {Titum}}, \bibinfo {author} {\bibfnamefont {E.}~\bibnamefont {Berg}},
  \bibinfo {author} {\bibfnamefont {M.~S.}\ \bibnamefont {Rudner}}, \bibinfo
  {author} {\bibfnamefont {G.}~\bibnamefont {Refael}}, \ and\ \bibinfo {author}
  {\bibfnamefont {N.~H.}\ \bibnamefont {Lindner}},\ }\bibfield  {title}
  {\enquote {\bibinfo {title} {Anomalous {Floquet}-{Anderson} {Insulator} as a
  {Nonadiabatic} {Quantized} {Charge} {Pump}},}\ }\href {\doibase
  10.1103/PhysRevX.6.021013} {\bibfield  {journal} {\bibinfo  {journal} {Phys.
  Rev. X}\ }\textbf {\bibinfo {volume} {6}},\ \bibinfo {pages} {021013}
  (\bibinfo {year} {2016})}\BibitemShut {NoStop}%
\bibitem [{\citenamefont {Gopalakrishnan}\ \emph {et~al.}(2011)\citenamefont
  {Gopalakrishnan}, \citenamefont {Lamacraft},\ and\ \citenamefont
  {Goldbart}}]{gopalakrishnan_universal_2011}%
  \BibitemOpen
  \bibfield  {author} {\bibinfo {author} {\bibfnamefont {S.}~\bibnamefont
  {Gopalakrishnan}}, \bibinfo {author} {\bibfnamefont {A.}~\bibnamefont
  {Lamacraft}}, \ and\ \bibinfo {author} {\bibfnamefont {P.~M.}\ \bibnamefont
  {Goldbart}},\ }\bibfield  {title} {\enquote {\bibinfo {title} {Universal
  phase structure of dilute {Bose} gases with {Rashba} spin-orbit coupling},}\
  }\href {\doibase 10.1103/PhysRevA.84.061604} {\bibfield  {journal} {\bibinfo
  {journal} {Phys. Rev. A}\ }\textbf {\bibinfo {volume} {84}},\ \bibinfo
  {pages} {061604} (\bibinfo {year} {2011})}\BibitemShut {NoStop}%
\bibitem [{\citenamefont {Sedrakyan}\ \emph {et~al.}(2012)\citenamefont
  {Sedrakyan}, \citenamefont {Kamenev},\ and\ \citenamefont
  {Glazman}}]{sedrakyan_composite_2012}%
  \BibitemOpen
  \bibfield  {author} {\bibinfo {author} {\bibfnamefont {T.~A.}\ \bibnamefont
  {Sedrakyan}}, \bibinfo {author} {\bibfnamefont {A.}~\bibnamefont {Kamenev}},
  \ and\ \bibinfo {author} {\bibfnamefont {L.~I.}\ \bibnamefont {Glazman}},\
  }\bibfield  {title} {\enquote {\bibinfo {title} {Composite fermion state of
  spin-orbit-coupled bosons},}\ }\href {\doibase 10.1103/PhysRevA.86.063639}
  {\bibfield  {journal} {\bibinfo  {journal} {Phys. Rev. A}\ }\textbf {\bibinfo
  {volume} {86}},\ \bibinfo {pages} {063639} (\bibinfo {year}
  {2012})}\BibitemShut {NoStop}%
\bibitem [{\citenamefont {Sedrakyan}\ \emph {et~al.}(2015)\citenamefont
  {Sedrakyan}, \citenamefont {Galitski},\ and\ \citenamefont
  {Kamenev}}]{sedrakyan_statistical_2015}%
  \BibitemOpen
  \bibfield  {author} {\bibinfo {author} {\bibfnamefont {T.~A.}\ \bibnamefont
  {Sedrakyan}}, \bibinfo {author} {\bibfnamefont {V.~M.}\ \bibnamefont
  {Galitski}}, \ and\ \bibinfo {author} {\bibfnamefont {A.}~\bibnamefont
  {Kamenev}},\ }\bibfield  {title} {\enquote {\bibinfo {title} {Statistical
  {Transmutation} in {Floquet} {Driven} {Optical} {Lattices}},}\ }\href
  {\doibase 10.1103/PhysRevLett.115.195301} {\bibfield  {journal} {\bibinfo
  {journal} {Phys. Rev. Lett.}\ }\textbf {\bibinfo {volume} {115}},\ \bibinfo
  {pages} {195301} (\bibinfo {year} {2015})}\BibitemShut {NoStop}%
\end{thebibliography}

\begin{thebibliography}{9}%
\makeatletter
\providecommand \@ifxundefined [1]{%
 \@ifx{#1\undefined}
}%
\providecommand \@ifnum [1]{%
 \ifnum #1\expandafter \@firstoftwo
 \else \expandafter \@secondoftwo
 \fi
}%
\providecommand \@ifx [1]{%
 \ifx #1\expandafter \@firstoftwo
 \else \expandafter \@secondoftwo
 \fi
}%
\providecommand \natexlab [1]{#1}%
\providecommand \enquote  [1]{``#1''}%
\providecommand \bibnamefont  [1]{#1}%
\providecommand \bibfnamefont [1]{#1}%
\providecommand \citenamefont [1]{#1}%
\providecommand \href@noop [0]{\@secondoftwo}%
\providecommand \href [0]{\begingroup \@sanitize@url \@href}%
\providecommand \@href[1]{\@@startlink{#1}\@@href}%
\providecommand \@@href[1]{\endgroup#1\@@endlink}%
\providecommand \@sanitize@url [0]{\catcode `\\12\catcode `\$12\catcode
  `\&12\catcode `\#12\catcode `\^12\catcode `\_12\catcode `\%12\relax}%
\providecommand \@@startlink[1]{}%
\providecommand \@@endlink[0]{}%
\providecommand \url  [0]{\begingroup\@sanitize@url \@url }%
\providecommand \@url [1]{\endgroup\@href {#1}{\urlprefix }}%
\providecommand \urlprefix  [0]{URL }%
\providecommand \Eprint [0]{\href }%
\providecommand \doibase [0]{http://dx.doi.org/}%
\providecommand \selectlanguage [0]{\@gobble}%
\providecommand \bibinfo  [0]{\@secondoftwo}%
\providecommand \bibfield  [0]{\@secondoftwo}%
\providecommand \translation [1]{[#1]}%
\providecommand \BibitemOpen [0]{}%
\providecommand \bibitemStop [0]{}%
\providecommand \bibitemNoStop [0]{.\EOS\space}%
\providecommand \EOS [0]{\spacefactor3000\relax}%
\providecommand \BibitemShut  [1]{\csname bibitem#1\endcsname}%
\let\auto@bib@innerbib\@empty
\bibitem [S1]{stringari_collective_1996}%
  \BibitemOpen
  \bibfield  {author} {\bibinfo {author} {\bibfnamefont {S.}~\bibnamefont
  {Stringari}},\ }\bibfield  {title} {\enquote {\bibinfo {title} {Collective
  {Excitations} of a {Trapped} {Bose}-{Condensed} {Gas}},}\ }\href {\doibase
  10.1103/PhysRevLett.77.2360} {\bibfield  {journal} {\bibinfo  {journal}
  {Phys. Rev. Lett.}\ }\textbf {\bibinfo {volume} {77}},\ \bibinfo {pages}
  {2360--2363} (\bibinfo {year} {1996})}\BibitemShut {NoStop}%
\bibitem [S2]{greiner_exploring_2001s}%
  \BibitemOpen
  \bibfield  {author} {\bibinfo {author} {\bibfnamefont {M.}~\bibnamefont
  {Greiner}}, \bibinfo {author} {\bibfnamefont {I.}~\bibnamefont {Bloch}},
  \bibinfo {author} {\bibfnamefont {O.}~\bibnamefont {Mandel}}, \bibinfo
  {author} {\bibfnamefont {T.~W.}\ \bibnamefont {Hänsch}}, \ and\ \bibinfo
  {author} {\bibfnamefont {T.}~\bibnamefont {Esslinger}},\ }\bibfield  {title}
  {\enquote {\bibinfo {title} {Exploring {Phase} {Coherence} in a {2D}
  {Lattice} of {Bose}-{Einstein} {Condensates}},}\ }\href {\doibase
  10.1103/PhysRevLett.87.160405} {\bibfield  {journal} {\bibinfo  {journal}
  {Phys. Rev. Lett.}\ }\textbf {\bibinfo {volume} {87}},\ \bibinfo {pages}
  {160405} (\bibinfo {year} {2001})}\BibitemShut {NoStop}%
\bibitem [S3]{rechtsman_photonic_2013s}%
  \BibitemOpen
  \bibfield  {author} {\bibinfo {author} {\bibfnamefont {M.~C.}\ \bibnamefont
  {Rechtsman}}, \bibinfo {author} {\bibfnamefont {J.~M.}\ \bibnamefont
  {Zeuner}}, \bibinfo {author} {\bibfnamefont {Y.}~\bibnamefont {Plotnik}},
  \bibinfo {author} {\bibfnamefont {Y.}~\bibnamefont {Lumer}}, \bibinfo
  {author} {\bibfnamefont {D.}~\bibnamefont {Podolsky}}, \bibinfo {author}
  {\bibfnamefont {F.}~\bibnamefont {Dreisow}}, \bibinfo {author} {\bibfnamefont
  {S.}~\bibnamefont {Nolte}}, \bibinfo {author} {\bibfnamefont
  {M.}~\bibnamefont {Segev}}, \ and\ \bibinfo {author} {\bibfnamefont
  {A.}~\bibnamefont {Szameit}},\ }\bibfield  {title} {\enquote {\bibinfo
  {title} {Photonic {Floquet} topological insulators},}\ }\href {\doibase
  10.1038/nature12066} {\bibfield  {journal} {\bibinfo  {journal} {Nature}\
  }\textbf {\bibinfo {volume} {496}},\ \bibinfo {pages} {196--200} (\bibinfo
  {year} {2013})}\BibitemShut {NoStop}%
\bibitem [S4]{jotzu_experimental_2014s}%
  \BibitemOpen
  \bibfield  {author} {\bibinfo {author} {\bibfnamefont {G.}~\bibnamefont
  {Jotzu}}, \bibinfo {author} {\bibfnamefont {M.}~\bibnamefont {Messer}},
  \bibinfo {author} {\bibfnamefont {R.}~\bibnamefont {Desbuquois}}, \bibinfo
  {author} {\bibfnamefont {M.}~\bibnamefont {Lebrat}}, \bibinfo {author}
  {\bibfnamefont {T.}~\bibnamefont {Uehlinger}}, \bibinfo {author}
  {\bibfnamefont {D.}~\bibnamefont {Greif}}, \ and\ \bibinfo {author}
  {\bibfnamefont {T.}~\bibnamefont {Esslinger}},\ }\bibfield  {title} {\enquote
  {\bibinfo {title} {Experimental realization of the topological {Haldane}
  model with ultracold fermions},}\ }\href {\doibase 10.1038/nature13915}
  {\bibfield  {journal} {\bibinfo  {journal} {Nature}\ }\textbf {\bibinfo
  {volume} {515}},\ \bibinfo {pages} {237--240} (\bibinfo {year}
  {2014})}\BibitemShut {NoStop}%
\bibitem [S5]{fukui_chern_2005}%
  \BibitemOpen
  \bibfield  {author} {\bibinfo {author} {\bibfnamefont {T.}~\bibnamefont
  {Fukui}}, \bibinfo {author} {\bibfnamefont {Y.}~\bibnamefont {Hatsugai}}, \
  and\ \bibinfo {author} {\bibfnamefont {H.}~\bibnamefont {Suzuki}},\
  }\bibfield  {title} {\enquote {\bibinfo {title} {Chern {Numbers} in
  {Discretized} {Brillouin} {Zone}: {Efficient} {Method} of {Computing}
  ({Spin}) {Hall} {Conductances}},}\ }\href {\doibase 10.1143/JPSJ.74.1674}
  {\bibfield  {journal} {\bibinfo  {journal} {J. Phys. Soc. Jpn.}\ }\textbf
  {\bibinfo {volume} {74}},\ \bibinfo {pages} {1674--1677} (\bibinfo {year}
  {2005})}\BibitemShut {NoStop}%
\bibitem [S6]{bellissard_change_1995s}%
  \BibitemOpen
  \bibfield  {author} {\bibinfo {author} {\bibfnamefont {J.}~\bibnamefont
  {Bellissard}},\ }\bibfield  {title} {\enquote {\bibinfo {title} {Change of
  the {Chern} number at band crossings},}\ }\href
  {http://arxiv.org/abs/cond-mat/9504030} {\bibfield  {journal} {\bibinfo
  {journal} {arXiv}\ }\textbf {\bibinfo {volume} {cond-mat/9504030}} (\bibinfo
  {year} {1995})}\BibitemShut {NoStop}%
\bibitem [S7]{young_dirac_2012}%
  \BibitemOpen
  \bibfield  {author} {\bibinfo {author} {\bibfnamefont {S.~M.}\ \bibnamefont
  {Young}}, \bibinfo {author} {\bibfnamefont {S.}~\bibnamefont {Zaheer}},
  \bibinfo {author} {\bibfnamefont {J.~C.~Y.}\ \bibnamefont {Teo}}, \bibinfo
  {author} {\bibfnamefont {C.~L.}\ \bibnamefont {Kane}}, \bibinfo {author}
  {\bibfnamefont {E.~J.}\ \bibnamefont {Mele}}, \ and\ \bibinfo {author}
  {\bibfnamefont {A.~M.}\ \bibnamefont {Rappe}},\ }\bibfield  {title} {\enquote
  {\bibinfo {title} {Dirac {Semimetal} in {Three} {Dimensions}},}\ }\href
  {\doibase 10.1103/PhysRevLett.108.140405} {\bibfield  {journal} {\bibinfo
  {journal} {Phys. Rev. Lett.}\ }\textbf {\bibinfo {volume} {108}},\ \bibinfo
  {pages} {140405} (\bibinfo {year} {2012})}\BibitemShut {NoStop}%
\bibitem [S8]{armitage_weyl_2018}%
  \BibitemOpen
  \bibfield  {author} {\bibinfo {author} {\bibfnamefont {N.~P.}\ \bibnamefont
  {Armitage}}, \bibinfo {author} {\bibfnamefont {E.~J.}\ \bibnamefont {Mele}},
  \ and\ \bibinfo {author} {\bibfnamefont {A.}~\bibnamefont {Vishwanath}},\
  }\bibfield  {title} {\enquote {\bibinfo {title} {Weyl and {Dirac} semimetals
  in three-dimensional solids},}\ }\href {\doibase
  10.1103/RevModPhys.90.015001} {\bibfield  {journal} {\bibinfo  {journal}
  {Rev. Mod. Phys.}\ }\textbf {\bibinfo {volume} {90}},\ \bibinfo {pages}
  {015001} (\bibinfo {year} {2018})}\BibitemShut {NoStop}%
\bibitem [S9]{aidelsburger_measuring_2015s}%
  \BibitemOpen
  \bibfield  {author} {\bibinfo {author} {\bibfnamefont {M.}~\bibnamefont
  {Aidelsburger}}, \bibinfo {author} {\bibfnamefont {M.}~\bibnamefont {Lohse}},
  \bibinfo {author} {\bibfnamefont {C.}~\bibnamefont {Schweizer}}, \bibinfo
  {author} {\bibfnamefont {M.}~\bibnamefont {Atala}}, \bibinfo {author}
  {\bibfnamefont {J.~T.}\ \bibnamefont {Barreiro}}, \bibinfo {author}
  {\bibfnamefont {S.}~\bibnamefont {Nascimbène}}, \bibinfo {author}
  {\bibfnamefont {N.~R.}\ \bibnamefont {Cooper}}, \bibinfo {author}
  {\bibfnamefont {I.}~\bibnamefont {Bloch}}, \ and\ \bibinfo {author}
  {\bibfnamefont {N.}~\bibnamefont {Goldman}},\ }\bibfield  {title} {\enquote
  {\bibinfo {title} {Measuring the {Chern} number of {Hofstadter} bands with
  ultracold bosonic atoms},}\ }\href {\doibase 10.1038/nphys3171} {\bibfield
  {journal} {\bibinfo  {journal} {Nature Phys.}\ }\textbf {\bibinfo {volume}
  {11}},\ \bibinfo {pages} {162--166} (\bibinfo {year} {2015})}\BibitemShut
  {NoStop}%
\end{thebibliography}

%

\cleardoublepage

\section*{Supplementary Information}

\renewcommand{\thefigure}{S\arabic{figure}}
\renewcommand{\theHfigure}{S\arabic{figure}}
 \setcounter{figure}{0}
\renewcommand{\theequation}{S.\arabic{equation}}
 \setcounter{equation}{0}
 \renewcommand{\thesection}{S\arabic{section}}
\setcounter{section}{0}

Here, we present calibration measurements and additional data (\ref{Sec:calib}), the detailed theoretical model (\ref{Sec:calc}), the calculation of the edge states in a tight-binding model (\ref{Sec:Edge_modes}) and the connection between the topological charge and the Berry curvature (\ref{Sec:theory}).

\section{Calibrations and additional measurements}
\label{Sec:calib}

\subsection{Influence of the harmonic trap}\label{SubSec_SI_trap}

\begin{figure}[!htb]
\includegraphics{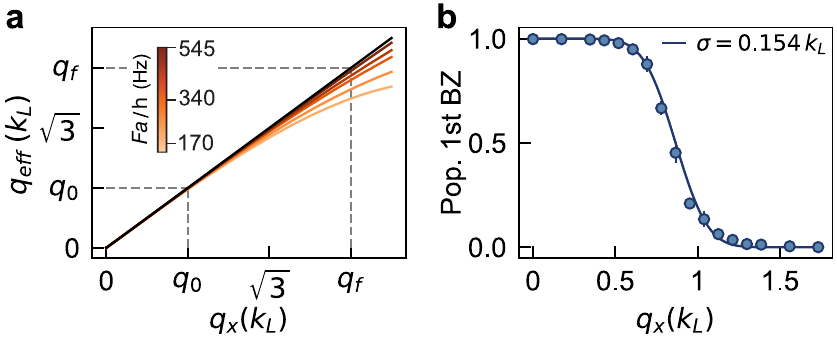}
\vspace{-0.cm} \caption{Calculated longitudinal quasimomentum center of mass (CoM) and calibration of the momentum space width. \textbf{a}. Calculated quasimomentum CoM $q_{\text{eff}}$ along the direction of the force as a function of the programmed quasimomentum $q_x$ in the presence of the harmonic trap and a static lattice with $V=6 \, E_r$ (see Sec.~\ref{Sec:calc}). 
The force is applied along the $\Gamma$-direction and varied between $Fa/h = 170  \, \text{Hz}$ and $Fa/h = 545  \,\text{Hz}$, as indicated by the colorbar, and $\sigma = 0.139 \, k_L$. 
The black line is the solution without the harmonic trap. 
The dashed lines mark the quasimomentum $q_0 = 0.5 \sqrt{3} \, k_L$ up to which we accelerate during the parameter ramp-up in many cases, and the final quasimomentum $q_f = 1.5 \sqrt{3} \, k_L$. 
\textbf{b}. Measurement of the momentum space width by observing the population transfer when driving adiabatically across the border of the first BZ. 
Each point is an average over five individual experimental realizations, errorbars indicate the standard error.
The solid line is an errorfunction fitted to the data to determine the Gaussian width $\sigma$.}
\label{Fig_S1}
\end{figure}

We apply a force on the cloud by accelerating the lattice, which leads to a longitudinal velocity of the atoms in the lab frame. 
Detuning the frequency of one laser beam by $\Delta f = \Delta \omega/(2\pi)$ changes the quasimomentum of the atoms by 
\begin{align}
\Delta q = \frac{2 \lambda_L m_{\text{K}} \Delta f}{3 \hbar}.
\label{Eq_LatAcc}
\end{align}
Changing the laser frequency linearly for a time $\Delta t$ gives rise to the force:
\begin{align}
F = \frac{\hbar\Delta q}{\Delta t} = m_{\text{K}} a_L.
\end{align}
The force is varied by changing the time $\Delta t$ and keeping the final detuning fixed. 
For the bandgap measurements the applied forces are large and $\Delta t$ is small, leading only to minor displacements in real space, so in this case the effect of the harmonic trap can be neglected. 
The transverse deflections were probed with smaller forces to ensure that we adiabatically move within a single band, yielding real-space displacements up to $\approx 100  \, \mu \text{m}$. 
In the presence of the harmonic trap the semiclassical equations of motion read:
\begin{align}
\dot{x} &= \frac{1}{\hbar}\frac{\partial \varepsilon}{\partial q_x}(\mathbf{q}) - \frac{1}{\hbar}\left(F_y - \frac{\partial V_{\text{trap}}}{\partial y} \right) \Omega(\mathbf{q}) + \frac{F_x}{m_{\text{K}}}t \notag \\ 
\dot{y} &= \frac{1}{\hbar}\frac{\partial \varepsilon}{\partial q_y}(\mathbf{q}) + \frac{1}{\hbar}\left(F_x - \frac{\partial V_{\text{trap}}}{\partial x} \right) \Omega(\mathbf{q}) + \frac{F_y}{m_{\text{K}}}t \notag \\
\dot{q}_x &= \frac{1}{\hbar}\left(F_x - \frac{\partial V_{\text{trap}}}{\partial x} \right) \notag \\
\dot{q}_y &= \frac{1}{\hbar}\left(F_y - \frac{\partial V_{\text{trap}}}{\partial y} \right),
\label{Eq_EoM}
\end{align}
where the trapping potential is given by $V_{\text{trap}} = 0.5 m_{\text{K}} \omega_r^2 (x^2 + y^2)$ with the mean trapping frequency in the $xy$-plane being $\omega_r = 2\pi \times 27.0 (4)\, \text{Hz}$ (see below).
The additional acceleration terms arise due to the motion of the lattice potential when transforming back into the lab frame and give rise to the longitudinal displacements mentioned above.
In these cases the restoring force of the harmonic trap becomes significant along the direction of the force leading to a reduction of the longitudinal displacement and quasimomentum. 
Hence, a different amount of Berry curvature is traversed in reciprocal space, potentially changing the transverse deflection. 
In Fig.~\ref{Fig_S1}a, the calculated longitudinal quasimomenta are shown for different forces applied along the $\Gamma$-direction as a function of the quasimomentum $q_f$ set by the lattice acceleration according to Eq.~(\ref{Eq_LatAcc}). 
At $q_x = q_0 = 0.5 \sqrt{3} \, k_L$, the changes are minor, even for the smallest force of $Fa/h = 170 \, \text{Hz}$ that was used for the parameter scan in Fig.~3 in the main text. 
But at the final value of $q_f = 1.5 \sqrt{3} \, k_L$ the quasimomentum is reduced to $q_{\text{eff}} \approx 1.25 \sqrt{3}\, k_L$.

Along the transverse direction the real space displacements are small leading only to minor changes in the quasimomentum due to the trap. 
To calculate the transverse deflections we numerically solved the set of equations in~(\ref{Eq_EoM}) including the band dispersion and the harmonic trap. 
The resulting transverse quasimomentum components were $q_{\perp} \le 0.005 \, k_L$ for all modulation parameters used in this work, meaning that the transverse band derivative is negligible, since the paths in reciprocal space are still well directed along the high-symmetry lines of the lattice. 
Hence, it is justified that that the transverse deflection measured in the experiments is indeed proportional to the Berry curvature.

The trapping frequency was measured in the presence of a static lattice with $V=6\,E_r$ by observing the breathing mode of the BEC insitu after a quench of the in-plane harmonic confinement. 
We fitted a 2D Gaussian to the absorption images, with the principle axes directed along the propagation directions of the trapping beams in the $xy$-plane, to extract the oscillation of the real-space width.
In our system, the trapping frequency along the vertical direction is about $8$-times larger than the in-plane frequency.
According to~\cite{stringari_collective_1996}, the in-plane trapping frequency can thus be extracted from the frequency $f_b$ of the breathing mode as:

\begin{align*}
f=\sqrt{\frac{3}{10}} f_b.
\end{align*}

The corresponding trapping frequencies along the dipole axes were $f_X = 27.9(7) \, \text{Hz}$ and $f_Y = 26.8 (4) \, \text{Hz}$ giving the weighted average value of $f = 27.0 (4) \, \text{Hz}$ mentioned above.

\subsection{Momentum space width}\label{SubSec_SI_sigma}

Due to finite temperatures, harmonic confinement and on-site interactions, the BEC is broadened in reciprocal space, which we describe by a symmetric Gaussian momentum distribution with width $\sigma$. 
The width was determined experimentally by performing a knife-edge measurement in reciprocal space: 
The quasimomentum is changed adiabatically by one reciprocal lattice vector along the $\Gamma$-direction using a force of $Fa/h = 204 \, \text{Hz}$ and performing bandmapping~\cite{greiner_exploring_2001} at certain quasimomenta along the path. 
Since the velocity component imposed by the moving lattice is directed opposite to the Bloch oscillation, the atoms appear, when bandmapping, at the $\Gamma$-point within the first BZ. When some of the atoms reach the edge of the BZ they appear at the $\Gamma$-point in the next BZ, so we count the relative population in the first BZ (similar to the bandgap measurements) depending on the quasimomentum. The amount of atoms in the first BZ is given by the integral over the Gaussian distribution and hence described by an errorfunction. 
An exemplary measurement is depicted in Fig.~\ref{Fig_S1}b along with the resulting fit.
The width of the error function was obtained from the fit, all other parameters were fixed. 
For every measurement of the transverse deflections we determined the width in reciprocal space immediately before or after the measurement and used this to calculate the corresponding theory values (see Sec.~\ref{Subsec:DeflCalc}).  

\begin{figure}[t]
\includegraphics{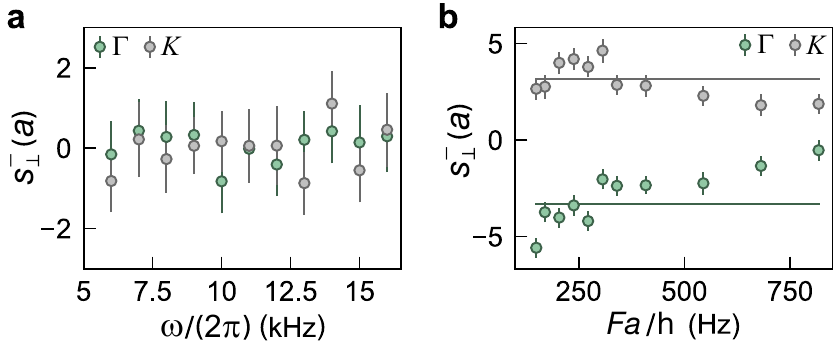}
\vspace{-0.cm} \caption{Transverse deflections $s_{\perp}^-$ in the Haldane and anomalous regime. \textbf{a}. Deflections vs. modulation frequency up to $q_0$ along the $\Gamma$- and $K$-directions with $F a/h = 204 \, \text{Hz}$ for $m=0.25$. Errorbars indicate the SEM.
\textbf{b}. Deflections in the anomalous regime ($m=0.24, \omega/(2\pi) = 10 \, \text{kHz}$) for $q_0 \rightarrow q_{\text{eff}} \approx 1.25 \sqrt{3} \, k_L$ depending on the applied force. 
The solid lines denote the corresponding theoretical values including the momentum space width. Errorbars indicate the SEM.}
\label{Fig_S2}
\end{figure}

\begin{figure*}[t]
\includegraphics{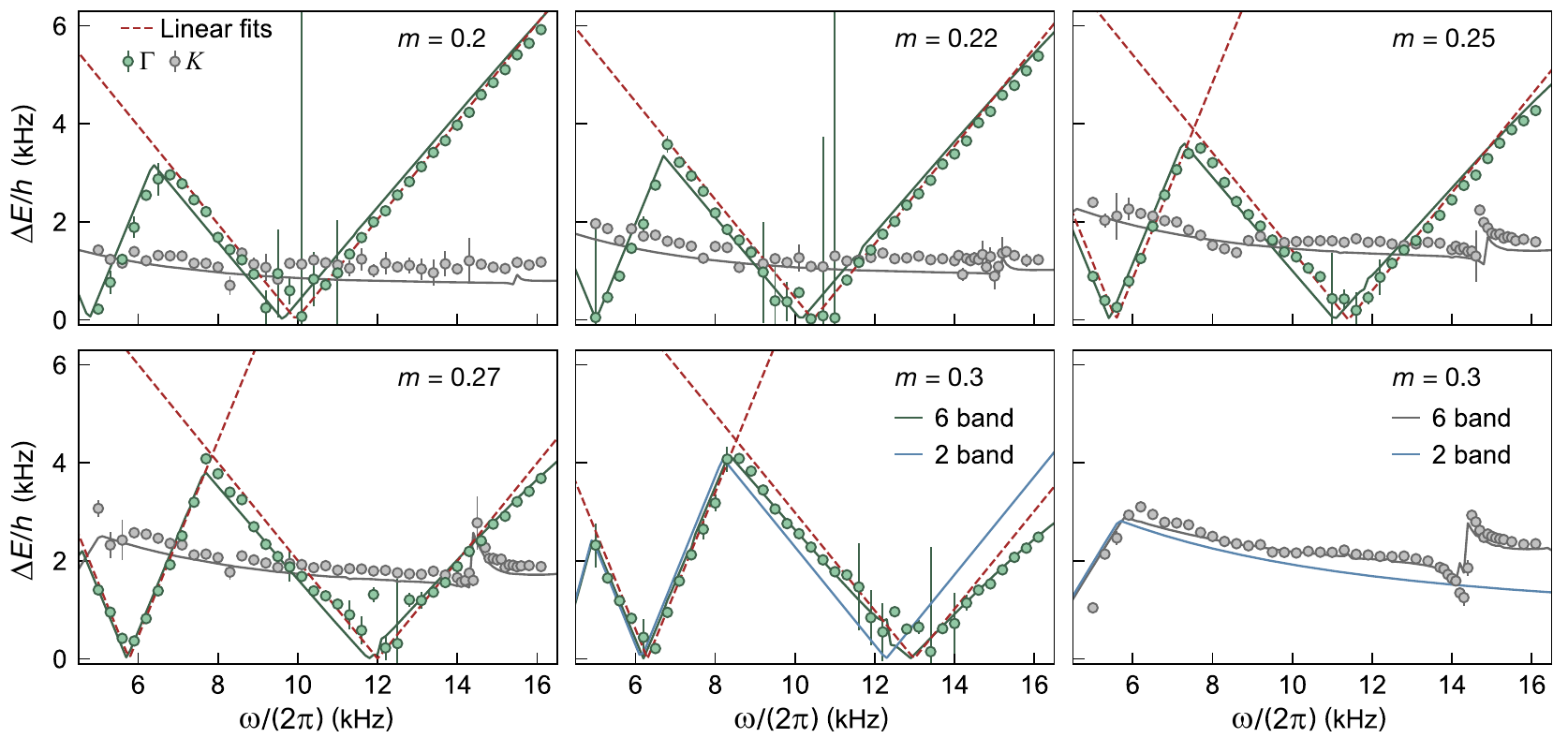}
\vspace{-0.cm} \caption{Energy gaps $\Delta E$ at $\Gamma$ and $K$ depending on the modulation frequency for modulation amplitudes $m=\{0.2, \, 0.22, \,0.25, \, 0.27, \, 0.3\}$ measured with St\"uckelberg interferometry using $Fa/h = 1360 \, \text{Hz}$.
The energy gap at K (gray circles) remains open over the full parameter range.
The gap at $\Gamma$ shows multiple closings, indicating the transitions from the Haldane to the anomalous and third phase with decreasing frequency.  
With increasing amplitude an avoided crossing appears at modulation frequencies around $\omega/(2 \pi)=15 \, \text{kHz}$.
The solid green and grey lines are the corresponding theoretical minimal gaps calculated using six bands, the solid blue lines in the last two panels are the theoretical gaps from a two-band model (see Sec.~\ref{Sec_SI_calc}).
The red dashed lines are fits $\propto |\omega|$ to the bandgaps at $\Gamma$ to determine the phase transitions (see text). 
The errorbars indicate fitting errors from the oscillation fits, every oscillation consists of 23 points each averaged over 3-4 individual experimental realizations.}
\label{Fig_S3}
\end{figure*}

\subsection{Deflections during ramp-up and test of the used forces} \label{Subsec_SI_rampup_forces}

To probe the Berry curvature in the Haldane and anomalous regime the modulation amplitude and partly the modulation frequency were ramped up while driving to $q_0$. 
Using the band structure calculations we verified that during the ramp-up the points where the two lowest bands potentially had touched and hybridized, which is the location of the additional negative Berry curvature in the anomalous phase, is always located away from the edge of the moving cloud in reciprocal space (see Methods). 
Since we are accelerating along high-symmetry lines in reciprocal space, the band derivatives along the transverse direction average to zero. 
Hence, there should be no deflection during the ramp-up and the measured transverse deflection can be assumed to correspond to a path in reciprocal space starting at $q_0$. 
This is confirmed by the data in Fig.~\ref{Fig_S2}a, showing the measured transverse deflections along the $\Gamma$- and $K$-direction up to a distance of $q_0$ for $m=0.25$ and different modulation frequencies.

We also verified that the forces we used to measure the transverse deflections were sufficiently small to avoid a reduction of the deflections due to excitations to the second band. 
We measured the deflections depending on the applied force when driving by $q_{\text{eff}} \approx 1.25 \sqrt{3} \, k_L$ along the $\Gamma$- and $K$-directions for modulation parameters in the anomalous regime (see Fig.~\ref{Fig_S2}b). 
The final quasimomenta for the lattice acceleration were chosen such that the effective length of the traversed path in reciprocal space was similar for all forces. 
For $Fa/h > 300 \, \text{Hz}$, the deflections along both directions are smaller than predicted by the theoretical calculations due to excitations to the second band. 
The modulation parameters chosen here lie close to the phase transition with energy gaps $\Delta E (K)/h = 1500 (30)\, \text{Hz}$ and $\Delta E(\Gamma)/h = 1110 (70)\, \text{Hz}$. 
The measured deflections saturate for smaller forces which happens earlier along the $K$-direction, also indicating the larger energy gap compared to $\Gamma$. 
In total, the chosen forces of $Fa/h = 170 \, \text{Hz}$ and $Fa/h = 204 \, \text{Hz}$ used for these modulation parameters are sufficiently small, which is also confirmed by the overall good agreement between the measured deflections and the theoretical calculations, where we assume population in a single band.

\subsection{Bandgap measurements for frequency scans}\label{SubSec_SI_bandgaps}

To explore the phase diagram shown in Fig.~1c we probed the bandgaps and Berry curvature for a broad range of modulation parameters in different topological regimes. 
The measured transverse deflections along the $\Gamma$- and $K$-directions are shown in the main text in Fig.~4a accompanied by the experimentally determined phase transitions. 
The corresponding gap measurements at $\Gamma$ and $K$ are displayed in Fig.~\ref{Fig_S3} together with the theoretical values from our model including the six lowest energy bands. 
For $m=0.3$ we also show the result of a model truncated to the lowest two energy bands.
At the phase transitions, the (absolute) energy gap at $\Gamma$ closes and reopens linearly with the modulation frequency for constant modulation amplitude. 
To determine the phase transition points, we fitted $\Delta E/h=n\cdot|\omega-\omega_0|/(2 \pi)$ to the slope on the left and right of the gap closings, with $n=1$ and $n=2$ for the first and second phase transition. 
The second phase transition could only be obtained for $m\geq 0.25$. 
The errors for the phase transitions are $\sigma_{\text{tot}}=\sqrt{\sigma_{\text{fit}}^2+\sigma_{\text{sys}}^2}$ with the fit errors $\sigma_{\text{fit}}$ and the systematic errors $\sigma_{\text{sys}}$. 
The latter are given by the step size $\Delta  \omega/(2\pi) = 300 \, \text{Hz}$ used in the energy gap measurements which is dominating the fit errors $\sigma_{\text{fit}} \in [20,70] \, \text{Hz}$.

We also measured the energy gaps at $K$ to validate our theoretical calculations and pick the forces for the deflection measurements appropriately. 
For large modulation frequencies and amplitudes the influence of the $p$-bands becomes significant which can be seen in the jumps of the energy gap around $\omega/(2\pi) = 15 \, \text{kHz}$: 
Due to the coupling between the different bands, gaps open at avoided crossings, which increase with modulation amplitude. 
These manifest in discontinuities in the effective Floquet bands and the corresponding energy gaps. 
The experimental data is well reproduced by a six-band model, signaling that coupling to even higher bands with $\mu > 6$ can be neglected. 
A two-band model fails to capture all signatures of the experimental data, as illustrated by the comparison for $m=0.3$ in Fig.~\ref{Fig_S3}:
The overall shape of the energy gap at $\Gamma$ is similar, but the first phase transition is shifted, whereas the difference between the two models decreases for smaller modulation frequencies. This is expected since the modulation frequency becomes more detuned from the energy gap to the $p$-bands. 
At $K$, the theoretical curves also coincide at small frequencies but in the Haldane regime the deviations are larger, especially the jumps at the avoided crossings are not captured by a two-band model. 
The general mechanism of the phase transitions is captured by a simple two-band model but to quantitatively describe the experiments performed here, a six-band model is necessary.

\subsection{St\"uckelberg interferometry}\label{SubSec_SI_stueckelberg}

\begin{figure}[t]
\includegraphics{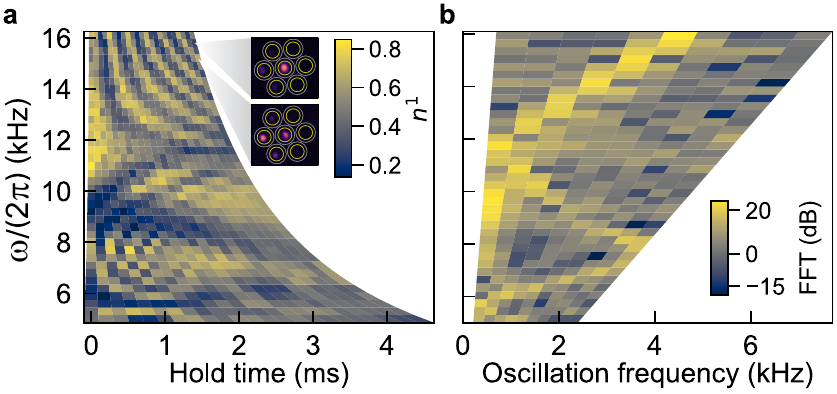}
\vspace{-0.cm} \caption{Raw data used to obtain the energy gaps. \textbf{a}. Relative population $n^1$ in the lowest band measured at $\Gamma$ depending on the hold time for different modulation frequencies and $m=0.25$. 
The reduction of the energy gap and thus oscillation frequency is clearly visible  illustrating the gap opening and closing.
Each population is an average over 3-4 individual experimental realizations. 
The two insets show raw images for $\omega/(2 \pi)=15.8 \, \text{kHz}$ and $\omega/(2 \pi)=15.2 \, \text{kHz}$ after a hold time of $22$ modulation cycles, brighter color indicates higher optical depth.
The yellow and grey circles indicate the areas used to define the pixel counts in each band and the corresponding background (see text). 
\textbf{b}. Fast Fourier transform (FFT) of the signal in \textbf{a} as a function of the St\"uckelberg oscillation frequency.}
\label{Fig_S4}
\end{figure}

All energy gaps presented in this work were measured using St\"uckelberg interferometry as described in the Methods. 
To quantify the amount of atoms in the first and second band, we take absorption images after performing bandmapping at $\Gamma$ giving distinct peaks corresponding to the different bands, as shown in the insets of Fig.~\ref{Fig_S4}a. 
The atoms in the lowest band appear in the center, whereas atoms in the second to sixth band are distributed over the outer peaks. 
The forces were chosen sufficiently large to ensure population of the second band but not too large to avoid excitations to the $p$-bands which can also be assumed to be small due to the good agreement of the measured energy gaps with the calculated minimal gaps. 
We sum up the pixels inside each of the seven regions of interest (ROIs) drawn as yellow circles with radius $R$ in the insets of Fig.~\ref{Fig_S4}a. 
To account for inhomogeneities in the background due to the finite size of the imaging beam, we also count the pixels in a larger ROI with radius $\sqrt{2} R$ (grey circles). 
The pixel counts for each peak are then obtained as $2 \Sigma_{R}- \Sigma_{\sqrt{2} R}$ and the relative population in the lowest band is given by the counts in the central peak divided by the total counts. 

An example of the population oscillations at $\Gamma$ for different modulation frequencies and $m=0.25$ is presented in Fig.~\ref{Fig_S4}a, already showing the decrease of the oscillation frequency towards the phase transitions. 
To obtain the points in Fig.~2 and Fig.~\ref{Fig_S3} we fit a sum of cosines to each population curve as described in the Methods section of the main text. 
However, the change in the oscillation frequency can also be seen directly by  performing a Fast Fourier transform (FFT) of the population oscillation (Fig.~\ref{Fig_S4}b) where the gap closings at the two phase transitions are clearly visible as well as additional small frequency components appearing around $\omega/(2 \pi) = 10 \, \text{kHz}$ probably arising from weak coupling to Floquet copies of the $p$-bands.

\subsection{Lifetimes}\label{SubSec_SI_lifetimes}

We measured the lifetime of the BEC at $\Gamma$ in all three topological regimes probed in this work. 
As described in the main text, in the anomalous phase the first band of the static lattice is adiabatically connected to the second band of the modulated lattice which has an energy minimum at $\Gamma$. 
Hence we probed the lifetime for the anomalous regime in the second band by ramping the modulation frequency and amplitude simultaneously in a non-linear fashion (see Methods) to directly access the anomalous regime. 
The third regime was probed in the first band, using a similar ramp-up but starting at a smaller modulation frequency. 
After ramping up the modulation we held the atoms at the $\Gamma$-point in the modulated lattice for different times $t=nT$ with $n \in \mathbb{N}$, then ramped down the modulation and performed bandmapping after $10 \, \text{ms}$ TOF. 
In the Haldane regime (for amplitude and phase modulation) the ramp time was fixed to $5 T$, in the anomalous and third regime we used $12 T$ and $7 T$-$8 T$ respectively, corresponding to $\approx 2 \, \text{ms}$. 
The population in the lowest band was then counted as $\Sigma^1 = 2 \Sigma_{R}- \Sigma_{\sqrt{2} R}$ for the central peak using the main and background ROIs described above for the St\"uckelberg oscillations. 
The population exhibited an exponential decay as a function of the hold time for most modulation parameters, so we fitted the function $\Sigma^1 (t) = A \text{e}^{-t/\tau} + y_0$ to it and extracted the parameters $A, y_0$ and the lifetime $\tau$, whereas all of them were constrained to be real and positive. 
The fitted offset was negligible in most cases, since we measured up to times $t$ at which almost no atoms were left. 

\begin{figure}[t]
\includegraphics{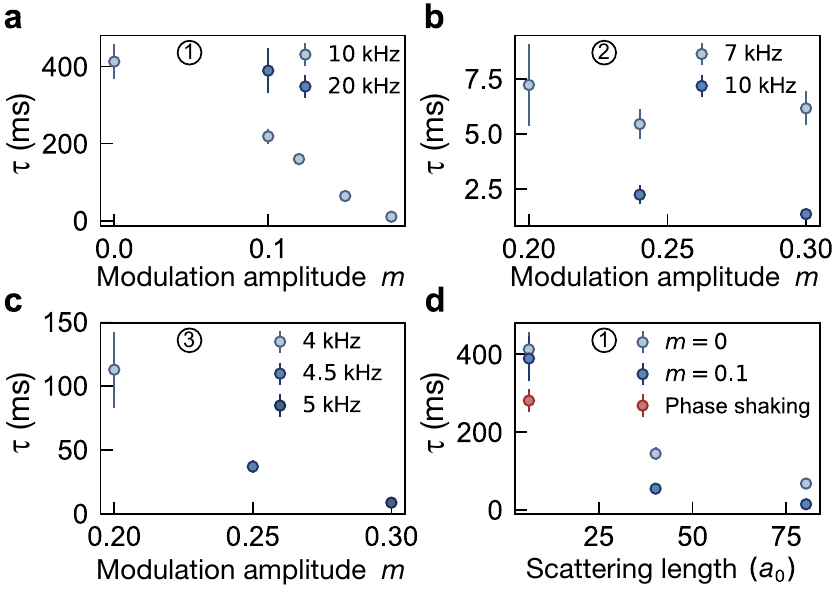}
\vspace{-0.cm} \caption{Measured lifetimes $\tau$ at $\Gamma$ vs. modulation parameters. 
\textbf{a}. Haldane regime in the first band for $a_s= 6.35\, a_0$. 
\textbf{b}. Anomalous regime in the second band for $a_s= 6.35\, a_0$. 
\textbf{c}. Third regime in the first band for $a_s= 6.35\, a_0$ whereas the modulation amplitude and frequency were chosen such that the gap at $\Gamma$ is similar. 
\textbf{d}. Haldane regime in the first band for $\omega/(2 \pi) = 20\, \text{kHz}$ vs. scattering length. 
The red data point is measured for phase modulation of the lattice with a frequency of $8 \, \text{kHz}$ and an amplitude of $6.6 \, \text{kHz}$ also realizing a Haldane system with the energy gap at $K$ calculated to be $\Delta E/h = 160 \,\text{Hz}$ similar to the measured value for $\omega/(2 \pi) = 20 \,\text{kHz}$ and $m=0.1$. 
Errorbars indicate fitting errors in all cases, every decay measurement consists of 23 points each averaged over 10 experimental realizations.}
\label{Fig_S5}
\end{figure}

In the Haldane regime (Fig.~\ref{Fig_S5}a) we compared the lifetime for different modulation amplitudes at $\omega/(2 \pi) = 10 \, \text{kHz}$ and for $m=0.1$ at $\omega/(2 \pi) = 20 \, \text{kHz}$. 
The lifetimes increase linearly for smaller modulation amplitudes and larger frequencies moving away from the first phase transition. 
The value for $m=0.1$ and $\omega/(2 \pi) = 20 \, \text{kHz}$ is comparable to the lifetime in the static lattice. 
In the anomalous regime (Fig.~\ref{Fig_S5}b) the lifetimes are much smaller and depend mainly on the modulation frequency. 
For $\omega/(2 \pi) = 7 \, \text{kHz}$ the system is deep in the anomalous regime and exhibits similar lifetimes for all amplitudes, whereas the lifetime is reduced significantly for $\omega/(2 \pi) = 10 \, \text{kHz}$ and slightly decreases with the modulation amplitude. 
In the third regime (Fig.~\ref{Fig_S5}c) the lifetimes increase again and strongly depend on the modulation amplitude and frequency. 
The parameters were chosen such that they have equal distance to the phase transition and the lifetimes are reduced by almost two orders of magnitude for larger amplitudes and higher frequencies. 
Based on the observed dependence of the lifetimes on the modulation parameters, we assume that these effects mainly originate from excitations to higher quasienergy bands, which are favored for larger modulation amplitudes and frequencies, increasing the coupling between the Floquet zones. In the Haldane regime, the lifetime increases for $f=20\, \text{kHz}$ compared tor $f=10\, \text{kHz}$, which can be understood as follows: At $f=20\, \text{kHz}$, $g^\pi(\Gamma) > g^0(\Gamma)$, reducing the coupling to the first Floquet copy of the second band, whereas the gap to the corresponding $p$-bands is still large. Increasing the modulation frequency further, reduces the lifetime again, since excitations to the $p$-bands are favored: For $f=30\, \text{kHz}$ and $m=0.1$ the measured lifetime is similar as for $f=10\, \text{kHz}$ (not shown in the plot).

The last panel of Fig.~\ref{Fig_S5} shows the lifetimes in the Haldane regime as a function of the scattering length which we can tune using a Feshbach resonance (see main text). 
All measurements so far were performed at $a_s = 6.35 \, a_0$. 
Increasing the on-site interaction considerably reduces the lifetimes in the static lattice, but even more in the modulated case, where the minimal lifetime is $\tau \approx 15 \, \text{ms}$ for $a_s = 80.25 \, a_0$. This suggests that there are also two-particle processes involved increasing the rate of excitations to higher Floquet bands. 

Overall, the smallest lifetimes measured are on the order of $2 \, \text{ms}$ in the second band and the anomalous regime being comparable to the maximal duration of $\approx 6 \, \text{ms}$ used in the deflection measurements. 
During these experiments the influence of the atom loss on the insitu images was minor, allowing for proper determination of the CoM-position by Gaussian fits in all cases. 
In the bandgap measurements the atom loss and heating was visible in the absorption images at long hold times, leading to damping of the oscillations. 

We also compared the lifetimes in the Haldane regime to the case of a Haldane system realized by circular phase modulation of the lattice (red data point in Fig.~\ref{Fig_S5}d), similar to~\cite{rechtsman_photonic_2013, jotzu_experimental_2014}. 
Here, the lattice was shaken at a frequency of $\omega_M/(2 \pi) = 8 \, \text{kHz}$ with an amplitude $b_M/(2\pi) = 6.6 \, \text{kHz}$ leading to an energy gap of $\Delta E /h = 160 \, \text{Hz}$ at the $K$-points which is similar to the corresponding gap for $m=0.1$ and $\omega/(2 \pi) = 20 \, \text{kHz}$. 
The phase shaking leads to a reduced lifetime compared to the amplitude modulation, which is nevertheless large with respect to the experimental times used here.

\section{Numerical calculations}\label{Sec_SI_calc}
\label{Sec:calc}

\subsection{Effective Hamiltonian and energy bands}
\label{Subsec:Heff}
In the experiments, we directly probe the properties of the bulk from which we can deduce the topological winding numbers associated with the band gaps. These topological invariants determine the existence of chiral edge modes in the system (not measured in the experiment).
To obtain the bulk energy bands and corresponding Berry curvatures in the modulated lattice we numerically calculated the effective Hamiltonian $H_{\text{eff}}$, which is defined via the time-evolution operator $U(T)$ over one full period of the drive:

\begin{align}
H_{\text{eff}} = \frac{i \hbar}{T} \text{ln}(U(T)), \quad
U(T) = \mathcal{T}e^{-\frac{i}{\hbar} \int_0^T H(t) dt},
\label{Eq_Heff}
\end{align}
where $\mathcal{T}$ denotes time-ordering and $\text{ln}$ the matrix logarithm.
To numerically calculate $H_{\text{eff}}$, the time-dependent Hamiltonian is evaluated at $300$ discrete timesteps $t_l$ within one driving period. 
We set $T=1$ for the integration over one driving period to simplify the numerics. 
For each set of parameters $(\textbf{q},m)$ we calculated the instantaneous Hamiltonian $H(t_l,\textbf{q},m)$ at each timestep $t_l$ in the basis of plane waves and projected it to its six lowest eigenstates:

\begin{align*}
H_p^{*}(t_l,\textbf{q},m) = M^{\dagger}(t_l,\textbf{q},m) \cdot H(t_l,\textbf{q},m) \cdot M(t_l,\textbf{q},m),
\end{align*}
where the columns of the matrix $M$ are the eigenstates of $H$ corresponding to the six lowest eigenvalues and $\cdot$ denotes matrix multiplication. 
The resulting $6\times6$-matrices $H_p^{*}$ are then transferred to a common basis consisting of the six lowest eigenstates of $H(t=0,\textbf{q}=0,m=0)$, being the columns of the Matrix $M_0$. 
The basis change is done as:

\begin{align*}
H_p(t_l,\textbf{q},m) &= B(t_l,\textbf{q},m) \cdot H_p^{*}(t_l,\textbf{q},m) \cdot B^{-1}(t_l,\textbf{q},m), \\
B(t_l,\textbf{q},m) &= M_0^{\dagger} \cdot M(t_l,\textbf{q},m).
\end{align*}

The time-evolution operator is then calculated from the projected Hamiltonians at each timestep:

\begin{align}
U(T,\textbf{q},m,f) = \Pi_l e^{-\frac{i}{\hbar} H_p(t_l,\textbf{q},m) \frac{\Delta t}{f}},
\end{align}

with $f=\omega/(2 \pi)$, and the effective Hamiltonian (in units of $\hbar \omega$) is given by

\begin{align}
H_{\text{eff}}(\textbf{q},m,f) = \frac{i}{2\pi} \text{ln}(U(T,\textbf{q},m,f)).
\end{align}

Due to the periodic driving, the energies are not bounded any more and the band of $H_{\text{eff}}$ that is connected to the lowest band of the static Hamiltonian not necessarily appears as the lowest. 
In our case, we are interested in the two lowest bands, which are adiabatically connected to the two $s$-bands of the static lattice. 

To extract the two lowest bands, we scanned the quasimomentum across the first BZ, calculated the six eigenstates and eigenenergies of each $H_{\text{eff}}(\textbf{q},m,f)$ and determined which of the states had the maximal overlap with the first and second eigenstate from the last $\mathbf{q}$-step. 
For the initial step we considered the overlap with the first two unit vectors, being the eigenstates of the two lowest bands in the static lattice. 
The state overlap is defined as the fidelity $\mathcal{F}_{ij}$:

\begin{align}
\mathcal{F}_{ij} = |\bra{\phi(\mathbf{q}_i,m,f)}\ket{\phi(\mathbf{q}_j,m,f)}|^2.
\end{align}

Especially at high modulation frequencies and amplitudes, all six bands couple and many avoided crossings appear. 
In the vicinity of these points, the eigenstate-overlap decreases and there can be several states having an overlap of similar magnitude with the first or second state of the last step.
If the overlap with the previous eigenstate dropped below a certain threshold, we used the overlap with the unit vectors instead to avoid false attributions.
The threshold value depends on the modulation parameters, i.e., for low modulation frequencies it could be set to $0.5$, using the eigenstate-overlap mostly everywhere.
By checking the bands in the  first BZ as well as on a 1D-high-symmetry line ($\Gamma-M-K-\Gamma$), we determined the optimal limits for the fidelity for each band and set of modulation parameters.
The results for two bands shown in Fig.~\ref{Fig_S3} were obtained by the same procedure but projecting the instantaneous Hamiltonian at each time step to its two lowest eigenstates.

\subsection{Transverse deflections}
\label{Subsec:DeflCalc}
From the eigenstates of the two lowest bands we numerically calculate the Berry curvature according to Ref.~\cite{fukui_chern_2005} on a rhombic grid spanning the first BZ. 
For the numeric integration of Eq.~(\ref{Eq_EoM}) we interpolate the Berry curvature and the band derivatives on a large quadratic grid spanning several BZs to be able to simulate the full trajectory including the momentum space extent of the BEC. 
The stepsize of the quadratic grid is $dq \approx 0.0145 \, k_L$ which was the maximal value at which the resulting real space positions did not change when decreasing the stepsize further. 

For most modulation parameters in the experiment, we effectively probed the deflections starting at a distance of $q_0$ in reciprocal space after ramping up the modulation parameters.
The quasimomentum after the ramp-up drive remains as $q_{||} \approx q_0$ and $q_{\perp}=0$, and hence $s_{\perp}=0$ (see Sec.~\ref{Subsec_SI_rampup_forces}). 
The longitudinal offset in real space was calculated by solving Eq.~(\ref{Eq_EoM}) with $\Omega(\mathbf{q})=0$, starting at $\Gamma$ and applying the respective force along the $\Gamma$- and/or $K$-direction for a time $\Delta t$ corresponding to $q_0$. 
The integration was performed for about $7300$ initial points in quasimomentum space lying on a circle with radius $0.5 \, k_L$ around $\Gamma$ to each of which we assigned a Gaussian weight according to the normalized momentum distribution of the BEC with width $\sigma$. 
The CoM position and quasimomentum after the ramp-up were then given as the weighted average over the corresponding final values. 
Note that we used the band derivatives for the final modulation parameters here, which turned out to give similar results as directly simulating the ramp-up by using the band derivatives for the different modulation parameters taken in between.

\begin{figure*}[t]
\includegraphics{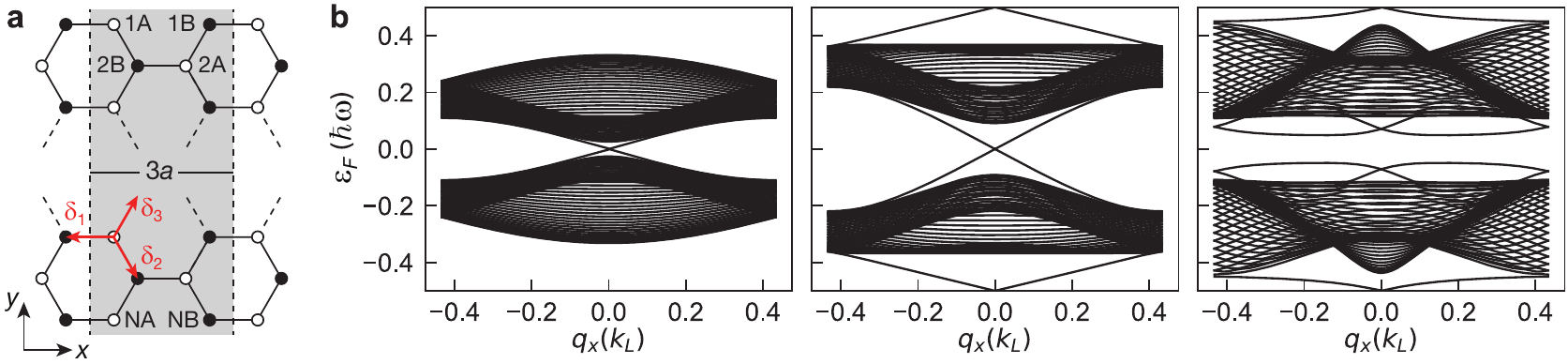}
\vspace{-0.cm} \caption{Schematic of the stripe-geometry and calculated quasienergy dispersions in the three topological phases plotted in the reduced zone scheme.
\textbf{a}. Schematic of the stripe geometry described in the text, terminated by an armchair-edge along $y$ and with periodic boundary conditions along $x$. The unit cell (gray shaded area) consists of $N$ dimers and has a width of $3a$, the red arrows denote the vectors $\delta_i$ connecting nearest neighbours.
\textbf{b}. Quasienergy dispersions from the two-band tight-binding model with $N=50$ for a modulation amplitude of $m=0.25$ and modulation frequencies $\omega/(2\pi)=\{ 16, 8, 4 \} \, \text{kHz}$ corresponding to the Haldane regime, anomalous regime and third regime hosting chiral edge modes in the $g^0$-gap, both gaps and the $g^\pi$-gap, respectively.}
\label{Fig_S6}
\end{figure*}

The deflections were then calculated by integrating Eq.~(\ref{Eq_EoM}) along the $\Gamma$- and $K$-direction with the initial points lying on a circle centered around the starting point after the ramp-up, given either by the values above or, when measuring along the $\Gamma$-direction in the first band and third regime or the second band and anomalous regime, by $q_{||}=0=q_{\perp}$ (see Methods). 
The time span was determined by the corresponding force and the programmed quasimomentum distance of $1.5 \sqrt{3} \, k_L - q_0$ (or $1.0 \sqrt{3} \, k_L$, respectively). 
The CoM deflection and quasimomentum were obtained as the average over the final values, again using the Gaussian weights of the independently calibrated density distribution. 
For modulation parameters lying in between the experimental points, the Gaussian width of the closest measured point was used.
The approximate final values $q_{\text{eff}}$ for the longitudinal quasimomentum mainly depend on the magnitude of the applied force and the time for which it is applied, whereas the influence of the band derivatives and the force direction is negligible.

\section{Calculation of edge states in a tight-binding model}
\label{Sec:Edge_modes}
As described in the main text, in a periodically-driven system the net number and chirality of edge modes per energy gap is given by the winding number of the respective gap, allowing for a full characterization of the system using topological invariants of the bulk only. In addition to the determination of the winding numbers we also calculated the energy spectrum of the effective Hamiltonian on a stripe-geometry displaying the dispersion of the edge states directly. We employed a two-band tight-binding model defined on a stripe terminated by an armchair-edge in the $y$-direction and with periodic boundary conditions along $x$ (Fig.~\ref{Fig_S6}a).

Due to hybridization between the $s$- and $p$-bands for large modulation frequencies, the lowest six energy bands have to be taken into account to quantitatively understand the position of the phase transitions for the model realized in our experiment (Sect.~\ref{Subsec:Heff}). Considering only the two lowest bands results in a shift of the transition points (see last two panels of Fig.~\ref{Fig_S3}), but the general nature of the topological phase diagram remains unchanged. In this section we present an approximate description based on a two-band tight-binding model with time-dependent nearest-neighbor hoppings that allows us to compute the dispersion of the edge modes directly.


The modulation of the relative intensities leads to a modulation of the distance between neighbouring lattice sites, which can be expressed as time-dependent tunneling matrix amplitudes in the tight-binding limit.
The unit cell of the stripe consists of $N$ dimers along the $y$-direction and has a width of $3a$ in the $x$-direction. Due to the periodicity along $x$, the Hamiltonian can be Fourier-transformed along this direction with quasimomentum $q_x \in [-\frac{\pi}{3a},\frac{\pi}{3a}]$. Including time-dependent nearest-neighbour tunneling along the directions $\mathbf{\delta_i}$ with amplitudes $J_i (t), \, i=\{1,2,3\}$ and setting the energy offset between the $A$- and $B$-sites to zero, the Hamiltonian reads

\begin{align}
\hat{H}_{\text{tb}}(q_x,t)&=-\sum\limits_{n, q_x} J_1(t)\left(e^{-iq_xa} \hat{\alpha}^\dagger_{q_x}(n)\hat{\beta}_{q_x}(n)+ c.c. \right)\notag \\  
&+ J_2(t)\left(e^{iq_x\frac{a}{2}} \hat{\alpha}^\dagger_{q_x}(n)\hat{\beta}_{q_x}(n+1) + c.c. \right)\notag \\ 
&+ J_3(t)\left(e^{iq_x\frac{a}{2}} \hat{\alpha}^\dagger_{q_x}(n+1)\hat{\beta}_{q_x}(n) + c.c. \right),
\label{Eq_Htb}
\end{align}
where $\hat{\alpha}^\dagger_{q_x}(n)$ and $\hat{\beta}^\dagger_{q_x}(n)$ create a particle with quasimomentum $q_x$ on the $n$th $A$- and $B$-site within the stripe.

To extract the time-dependent tunneling amplitudes, we fitted the energy bands of the two-band tight-binding model for the system without boundaries to the two lowest energy bands of the full Hamiltonian for a fixed modulation amplitude at every timestep $t_l$ within one driving period. The fit was performed on both energy bands in the entire $2$D-BZ, yielding the values of $J_i (t_l), \, i=\{1,2,3\}$ within one modulation cycle. In the full six-band Hamiltonian of our time-dependent honeycomb lattice model, particle-hole symmetry is broken resulting in an asymmetry of the two $s$-bands. This could be accounted for by including next-nearest-neighbour hoppings and coupling to $p$-orbitals, however, for a conceptual understanding of the phase diagram, the simple two-band model of Eq.~(\ref{Eq_Htb}) is sufficient.
In general, the nearest-neighbour hopping amplitude between two sites is expected to depend exponentially on the height of the potential barrier between the sites. Hence, we described the time-dependence of the hopping amplitudes as
\begin{align}
J_i(t) = A\, e^{B\, \text{cos}(\omega t + \phi_i)}+C \quad i=\{1,2,3\},
\end{align}
with $\phi_i=\frac{2\pi}{3}\times (i-1)$ and $A$, $B$ and $C$ are free variables that depend on the modulation amplitude. This function was fitted to the extracted hopping amplitudes. Using the time dependent hoppings we calculated the effective Hamiltonian by integration of $\hat{H}_{\text{tb}}(q_x,t)$ over one driving period according to Eq.~(\ref{Eq_Heff}) for every $q_x$.
The resulting $2N$ quasienergies are shown in Fig.~\ref{Fig_S6}b as a function of the quasimomentum $q_x$ for $N=50$, $m=0.25$ and different modulation frequencies describing the three topological phases. The first plot with $\omega/(2\pi)=16 \, \text{kHz}$ corresponds to the Haldane regime, where a pair of chiral edge modes is visible in the gap at zero quasienergy. For a system with an armchair edge being periodic along $x$, the $\Gamma$- and $K$-point are both displayed at $q_x=0$. At $\omega/(2\pi)=8 \, \text{kHz}$ the system is in the anomalous phase exhibiting an additional pair of edge modes in the $g^\pi$-gap between FBZs. In the third regime with $\omega/(2\pi)=4 \, \text{kHz}$ the are no edge modes at zero quasienergy in the $g^0$-gap, but there exist chiral edge modes in the $\pi$-gap, characterizing a Haldane-like topological system.

\section{Winding numbers and topological charge}
\label{Sec:theory}

\begin{figure}[t]
\includegraphics[]{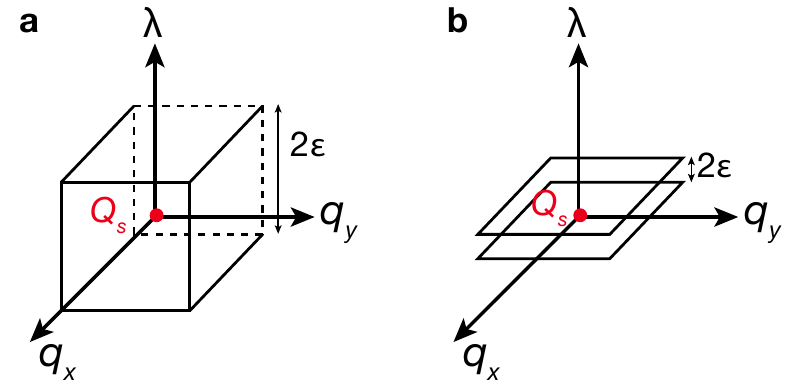}
\vspace{-0.cm} \caption{Schematic drawing of the topological charge $Q_s$ and the integration surfaces discussed below. 
\textbf{a}. Cube with height $2\varepsilon$ around $Q_s$ in the three-dimensional parameter space $(\mathbf{q},\lambda)$. 
\textbf{b}. Two two-dimensional surfaces in $\mathbf{q}$-space separated by $2\varepsilon$.}
\label{Fig:TCcube}
\end{figure}

The change in winding number across a topological phase transition is defined via the topological charge $Q^j_s$ of the band touching singularity as defined in Eq.~(1) in the main text. We consider a three-dimensional parameter space spanned by the quasimomentum $\mathbf{q}$ and $\lambda$, which smoothly connects a family of Hamiltonians $H(\mathbf{q},\lambda)$ (white arrows in Fig.~2a). For two-band models this Hamiltonian can be expressed as $H(\mathbf{q},\lambda)=\sum_{\alpha=x,y,z}h_\alpha(\mathbf{k}) \sigma_\alpha + h_0(\mathbf{k})\mathbb{1}$, where $\bm{\sigma}=(\sigma_x,\sigma_y,\sigma_z)$ is the vector of Pauli matrices and $\mathbb{1}$ is the identity matrix.
In generic cases we can make a Taylor expansion of the dispersion relation in the vicinity of the gap closing point~\cite{bellissard_change_1995} that occurs at $\xi:=(\mathbf{q}_s,\lambda_s)$. The resulting Hamiltonian can be expressed in the form of a Weyl Hamiltonian

\begin{equation}
H_s=v_{\beta\alpha} \xi_\beta \sigma_\alpha,
\label{eq:WeylHam}
\end{equation}
where $v$ is a $3\times 3$ matrix. The topological charge of the singularity is then given by~\cite{young_dirac_2012,armitage_weyl_2018} 

\begin{equation}
Q_s=\text{sgn}( \det(v)).
\end{equation} 
The Hamiltonian (\ref{eq:WeylHam}) has the form of a general three-parameter Hamiltonian $H_s=\mathbf{v}'\cdot \bm{\sigma}$, whose Berry curvature of the upper and lower state is described by magnetic monopoles $\bm{\Omega}^{\pm} = \mp \frac{1}{2}\frac{\mathbf{v}'}{v'^2}$ and there is a singularity at $\mathbf{v}_s'=0$. 
The Berry flux through a closed surface $\Sigma_c$ containing the singularity $\mathbf{v}_s'=0$ is given by
\begin{equation}
\phi^{\pm}=\mp \frac{1}{2}\int_{\Sigma_c} \text{d}\bm{\Sigma} \cdot \bm{\Omega}_{\pm} =  \mp 2\pi Q_s. 
\end{equation}
Note that here we defined the topological charge $Q_s$ by the Berry flux of the energy band below the respective energy gap. In Floquet systems there are two independent gaps and the topological charge $Q_s^0$ in the gap $g^0$ is defined via the Berry curvature of the lower band $\Omega^-$ and accordingly, the topological charge $Q_s^\pi$ in the gap $g^\pi$ is defined via the Berry curvature of the upper band $\Omega^+$.

One possibility to measure the topological charge associated with the singularity is to determine the flux through a sphere containing the band touching point or equivalently through a cube as depicted in Fig.~\ref{Fig:TCcube}a, which would require Berry flux measurements through the six surfaces of the cube in $(\mathbf{q},\lambda)$-space. This idea, however, can be simplified, if we consider the limit of $2\varepsilon\rightarrow 0$ and shift the origin trivially, such that the singularity located at $\xi$ is at the origin $\xi=0$. In this limit, the Berry flux through the two surfaces just before and just after the phase transition (Fig.~\ref{Fig:TCcube}b) in $\mathbf{q}$-space is determined by

\begin{align}
\phi^-_{\pm\varepsilon} &=\int_{-q_x^0}^{q_x^0} \text{d}q_x \int_{-q_y^0}^{q_y^0}\text{d}q_y \quad \Omega^-_{\lambda_s\pm\varepsilon}(\mathbf{q_s}) \nonumber \\
\phi^-_{\pm\varepsilon}& \quad \xrightarrow[]{\varepsilon \rightarrow 0} \quad \pm\pi 
\end{align}

Infinitesimally away from the band-touching singularity, the Berry curvature is perfectly localized in the $q_x q_y$-plane giving rise to a flux of $\pm\pi$.
Thus, in order to determine the sign of the topological charge we simply need to detect the sign of the $\pi$ Berry flux on both sides of the phase transition:
\begin{equation}
\frac{1}{2\pi} \Delta\phi^-(\mathbf{q}_s)\quad \xrightarrow[]{\varepsilon \rightarrow 0}\quad Q^0_s,
\end{equation}
with $\Delta\phi^-(\mathbf{q}_s)=\left(\phi^-_{+\varepsilon}-\phi^-_{-\varepsilon}\right)$.
The Berry flux $\phi$ is proportional to our measured deflections $s_\perp$, however, in the experiment we perform a weighted average according to the momentum distribution of our condensate as discussed in Section~\ref{Subsec:DeflCalc}. Nonetheless, if the spread is not too large and if we can perform the measurement close enough at the phase transition point, we can identify the topological charge of the singularity by determining the sign of the local Hall drifts across the phase transition~\cite{jotzu_experimental_2014,aidelsburger_measuring_2015}:

\begin{equation}
Q^0_s = \text{sgn}\left( \Delta s^-_\perp(\mathbf{q}_s )\right)=-\text{sgn}\left( \Delta s^+_\perp(\mathbf{q}_s )\right)
\end{equation}
Equivalently, the topological charge of the $\pi$-gap is determined by
\begin{equation}
Q^\pi_s = -\text{sgn}\left( \Delta s^-_\perp(\mathbf{q}_s )\right)=\text{sgn}\left( \Delta s^+_\perp(\mathbf{q}_s )\right) .
\end{equation}

\end{document}